\documentclass[journal,twoside,web]{ieeecolor}
\usepackage{generic}
\usepackage{cite}
\usepackage{amsmath,amssymb,amsfonts}
\usepackage{graphicx}
\usepackage{textcomp}
\usepackage{enumerate}
\usepackage{graphics}
\usepackage[font={small}]{caption}
\usepackage{multirow}
\usepackage{dsfont}
\usepackage{epsfig}
\usepackage{cite}
\usepackage{algorithm}
\usepackage[noend]{algpseudocode}

\def\R{\mathbb{R}}

\makeatletter
\newcommand*{\rom}[1]{\expandafter\@slowromancap\romannumeral #1@}
\makeatother

\newcommand{\E}{\mathbb{E}}
\newcommand{\PP}{\mathbb{P}}

\newcommand{\F}{\mathbb{F}}
\newcommand{\EE}{\mathcal{E}}

\newcommand{\bx}{{\bf x}}
\newcommand{\by}{{\bf y}}

\newcommand{\bz}{{\bf z}}

\def\mbb{\mathbb}
\def\mcal{\mathcal}

\newcommand{\mcx}{{\mathcal{X}}}
\newcommand{\mcy}{{\mathcal{Y}}}
\newcommand{\mcz}{{\mathcal{Z}}}
\newcommand{\mcd}{{\mathcal{D}}}

\newcommand{\qed}{\hfill \rule{1.2ex}{1.2ex} \\}
\newtheorem{theorem}{Theorem}[section]

\newtheorem{lemma}{Lemma}[section]
\newtheorem{proposition}{Proposition}[section]
\newtheorem{remark}{Remark}[section]

\newtheorem{definition}{Definition}[section]

\begin{document}

\title{The Defense of Networked Targets \\ in  General Lotto games}

\author{Adel Aghajan, Keith Paarporn, Jason R. Marden
\thanks{ A. Aghajan and J. R. Marden are with the Department of Electrical and Computer Engineering at the University of California, Santa Barbara, CA. K. Paarporn is with the Department of Computer Science at the University of Colorado, Colorado Springs. Contact: \texttt{ \{adelaa,jrmarden\}@ucsb.edu}, \texttt{kpaarpor@uccs.edu}. This work is supported by ONR grant \#N00014-20-1-2359, AFOSR grants \#FA9550-20-1-0054 and \#FA9550-21-1-0203, and the Army Research Lab through the ARL DCIST CRA \#W911NF-17-2-0181.}
}

% \date{}
\maketitle
% \thispagestyle{plain}
% \pagestyle{plain}
% \sloppy

\begin{abstract}
    Ensuring the security of networked systems is a significant problem, considering the susceptibility of modern infrastructures and technologies to adversarial interference. A central component of this problem is how defensive resources should be allocated to mitigate the severity of potential attacks on the system. In this paper, we consider this in the context of a General Lotto game, where a defender and attacker deploys resources on the nodes of a network, and the objective is to secure as many links as possible. The defender secures a link only if it out-competes the attacker on both of its associated nodes. For bipartite networks, we completely characterize equilibrium payoffs and strategies for both the defender and attacker. Surprisingly, the resulting payoffs are the same for any bipartite graph. On arbitrary network structures, we provide lower and upper bounds on the defender's max-min value. Notably, the equilibrium payoff from bipartite networks serves as the lower bound. These results suggest that more connected networks are easier to defend against attacks.  
    % We confirm these observations through numerical studies, where we show that tighter upper bounds can be calculated for arbitrary graphs. 
    We confirm these findings with simulations that compute deterministic allocation strategies on large random networks. This also highlights the importance of randomization in the equilibrium strategies.
    
    % We then highlight the value of randomized allocation strategies by comparing the performance of the defender when restricted to deterministic allocation strategies.
    
    % Indeed, we also establish that complete graphs (resp. bipartite) provide payoff guarantees for the widest (resp. smallest) range of opponent budgets when the defender is restricted to deterministic allocations.
\end{abstract}

\section{Introduction}

Networks are ingrained in modern technological society, thanks to advances in computing, communication, and control. Critical infrastructures, transportation networks, and cyber-physical systems are a few among many examples of systems that operate through complex interconnections. While their distributed nature gives rise to operations at unprecedented scale and efficiency, it also introduces vulnerabilities to adversarial interference.

A central component of ensuring the security of networked systems is the strategic allocation of limited resources to defend against potential attacks.  For example, allocating firewall and malware detectors, ensuring secure state estimation in critical infrastructures, and deploying defensive assets are among many problems requiring the strategic allocation of limited resources \cite{cardenas2009challenges,Pasqualetti_2013,Cortes_2004,alpcan2010network,tambe2011security,Pasqualetti_2013,Vamvoudakis_2014,amini2016dynamic,bai2017data,Milosevic_2020,Duan_2020,Ferdowsi_2020,Abdallah_2020,etesami2019dynamic,xing2021security}.

In this paper, we focus on the competitive allocation of resources between an attacker and a defender of a network. We formulate such a setting in the context of a General Lotto game. The General Lotto game is a popular variant of the famous Colonel Blotto game, wherein two opponents simultaneously allocate their limited resources against each other in order to secure multiple valuable battlefields \cite{Borel,Gross_1950,Roberson_2006,Schwartz_2014,Thomas_2018,Myerson_1993,Hart_2008,Kovenock_2021}. Colonel Blotto games and General Lotto games have been studied for well over 100 years, where its primary line of research has focused on characterizing equilibrium strategies and payoffs. More recently, they have been utilized to model complex adversarial environments that are relevant to many applications of interest \cite{Ferdowsi_2020,Guan_2019,shishika2022dynamic,paarporn2021division,paarporn2022asymmetric,aghajan2023extension}. 

In its classic formulation, a player's objective is to accumulate as much value as possible by securing individual battlefields. While this model is descriptive of many types of applications, this classic objective fails to precisely capture many other important scenarios. In particular, the operation of electricity grids, cyber networks, oil pipelines, and logistics chains requires uninterrupted interaction and communication between collaborating network nodes. In these applications, the defender's success in protecting such networked systems depends on securing certain \emph{subsets} of nodes, rather than securing as many individual nodes as possible. Ensuring the security of networked systems often requires the protection of certain graph characteristics, such as sub-networks or connected paths \cite{Shahrivar_2014,sandberg2015cyberphysical,Guan_2019,shishika2022dynamic,Ferguson_2022_Allerton,aghajan2023extension}.

% Characterizing equilibrium strategies in Colonel Blotto games, even in its classic formulation (no networks), is known to be extremely challenging, and has remained an open problem for over a century \cite{Borel,Gross_1950,Roberson_2006,Schwartz_2014,Thomas_2018}. This is primarily due to the fact that pure strategy equilibria do not exist for most interesting cases. The General Lotto game is a relaxed version where randomized strategies need only satisfy the resource budget in expectation \cite{Myerson_1993,Hart_2008,Kovenock_2021} instead of with probability one. It is more amenable to analysis while maintaining essential aspects of competitive resource allocation. Equilibrium characterizations of General Lotto games in the classic setting are well-established \cite{Hart_2008,Kovenock_2021}.

In this paper, we formulate a networked General Lotto game where a defender and attacker allocate resources to the nodes of a network. The defender's objective is to preserve the functioning of as many edges in the network as possible -- an edge is able to function if both of its endpoint nodes are under the control of the defender. Thus, in order to secure any given edge in the network, the defender is required to send more resources to \emph{both} endpoint nodes than the attacker. On the other hand, the attacker's objective is to disrupt the functioning of as many edges in the network as possible. As such, the attacker only needs to control \emph{at least one} of the endpoint nodes to disrupt the edge. The performance for each player is measured by the fraction of total edges in the network that it has secured. These objectives highlight the inherent asymmetry that exists in attack-defense scenarios. Namely, the asymmetry favors attackers, as attacks are more easily amplified by the network's connectivity properties (e.g. spreading of malware, disruption of communications, etc.).  

The networked setting vastly differs from the classical formulation, and consequently the resulting equilibrium strategies will be vastly different. To illustrate, consider the figure below: 

\begin{figure}
    \centering
    \includegraphics[scale=0.22]{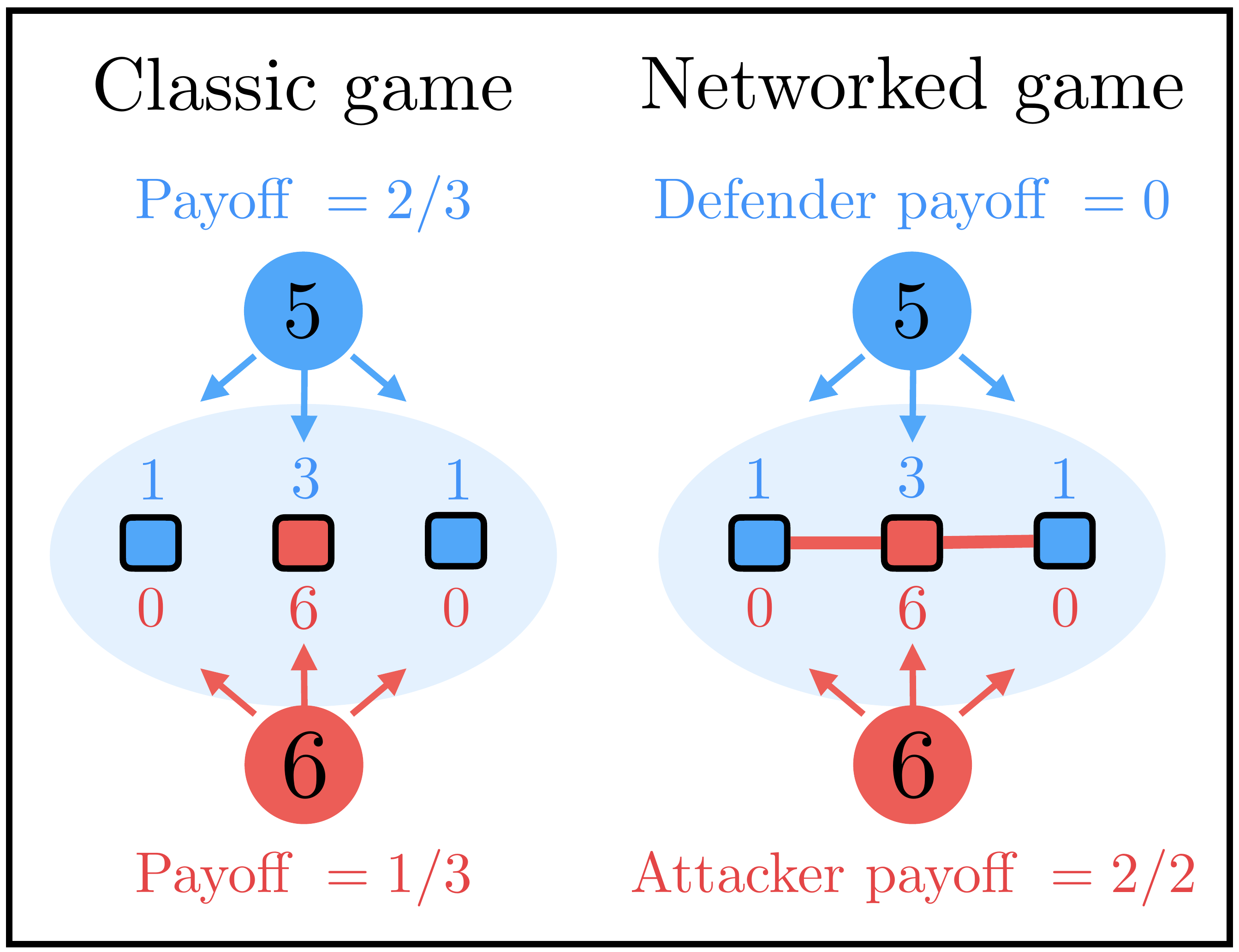}
    % \caption{}
    % \label{fig:first_example}
\end{figure}

There are three battlefields. In the classic setup, the objective is to win as many battlefields as possible, wherein each battlefield has a value of $1/3$. For the allocation strategies shown, the blue player (defender) wins two battlefields and thus obtains a payoff of $2/3$. However, for the same allocation strategies in the networked setup, the defender still wins both battlefields, but \emph{does not} attain any positive payoff. This is because it is necessary (but not sufficient) for the defender to win the center battlefield in order to secure any of the two edges (here, each are worth 1/2). 

Thus, one expects the equilibrium (equivalently, optimal) strategies in such settings to depend on the network structure under which the battlefields are arranged. Indeed, one of the main goals of this paper is to derive equilibrium strategies in networked General Lotto games of the above form. Moreover, a central question we seek to address in this paper is how the performance of the attacker and defender may change depending on characteristics of the network structure. We further illustrate the intricacies that can arise in simplified examples below.

\subsection{First illustrative examples}

To generate some initial intuition, we consider the following simplified setup. The defender $\mcx$ has a budget of $X \in \mathbb{R}_{\geq 0}$ resources, and the attacker $\mcy$ has $Y\in \mathbb{R}_{\geq 0}$ resources. Each must decide how to allocate their resources to the nodes of a network (suppose there are $n$ nodes). The set of feasible allocations for $\mcx$ is the set of vectors $\{(x_1,\ldots,x_n) \in \mathbb{R}_{\geq 0}^n : \sum_{i=1}^n x_i = X\}$ (and similarly for $\mcy$). Performance is measured as the fraction of edges that a player secures. In order to secure an edge of the network, $\mcx$ must win both endpoint nodes, whereas $\mcy$ only needs to win at least one. If they tie on a node, then we will assume the node is awarded to the attacker $\mcy$. Three different networks are shown in Figure \ref{fig:bipartite_examples}, where each player has a budget of 6 resources. Here, we illustrate that the graph structures significantly impact the players' attainable performances.

\vspace{1mm}

\noindent(a) Notice that in the star graph with six nodes (leftmost diagram), the attacker is guaranteed to secure the entire network by allocating resources only to the center node. This is because all edges are connected to the center. The defender here does not have enough resources to counter this strategy, and thus cannot secure a single edge. 

\vspace{1mm}

\noindent(b) Let us now consider the ring graph with six nodes (center diagram). For the same budgets, the attacker now cannot guarantee that it secures the entire network. This is because the attacker wins at least two edges by securing a single node, regardless of what the defender does.
% , and therefore must win three alternating nodes to secure the entire network. 
The best payoff it can guarantee itself on this network is two edges out of six. 

\vspace{1mm}

\noindent(c) A similar analysis applies to the line graph with five nodes (rightmost diagram). The attacker here can only guarantee that it secures two edges out of four. 
% There are multiple strategies that achieve this, one of which is shown. Another is where $\mcy$ allocates its entire budget of 6 to the center node. 

\vspace{1mm}

Clearly, the players' performance guarantees in these simplified examples are shaped by the structure of the networks. This paper focuses on analyzing the \emph{Network General Lotto} game (Section \ref{sec:formulation}), wherein randomized allocations that satisfy the budget in expectation are permitted. Surprisingly, our analysis finds that in the Network General Lotto game, the performance guarantees \emph{do not change} across the three networks shown in Figure \ref{fig:bipartite_examples}. That is, a player's performance in an equilibrium is \emph{identical} on the star, ring, and line network. A summary of our main contributions are given below.

% Indeed, analysis of the General Lotto game allows us to distinguish the effect of network structures on the equilibria of the game.

% The primary reason to study a General Lotto formulation is that it permits more analytic tractability \cite{Myerson_1993} over a Colonel Blotto formulation (randomized allocations must satisfy budget w.p. 1).

\subsection{Our Contributions}

Among the primary contributions of this paper, we establish equilibrium payoffs and strategies for the Network General Lotto game in the class of bipartite networks (Theorem \ref{thm:NL2}), which includes star graphs, rings with an even number of nodes, line graphs, tree graphs, and many others. Despite the variety of topologies that belong to the class of bipartite networks, the equilibrium payoffs for any bipartite network are \emph{identical}, and depends only on the relative budgets between the players. Hence, the payoffs are independent of any other characteristics. However, the equilibrium allocation strategies are intimately linked to the specific network topology.

Beyond bipartite networks, we identify analytical lower and upper bounds on the defender's security payoff guarantee on any network (Theorem \ref{thm:BoundsGeneral}). The lower bound coincides with the equilibrium payoff attained on any bipartite network. This suggests that bipartite networks are the ``easiest" networks to attack and ``hardest" to defend. The defender can guarantee the lower bound payoff by implementing its equilibrium strategy characterized for bipartite networks. 

% The upper bound in Theorem \ref{thm:BoundsGeneral} on the defender's payoff is derived by a particular attacker strategy, where the attacker randomizes its allocation strategy over a set of vertex covers. 

% Through numerical means, we are able to calculate tighter upper bounds by solving convex optimization problems. Our numerical studies reinforce the conclusion that the defender performs well on graphs with higher edge density, with complete graphs on the high extreme and bipartite graphs on the low extreme.

\begin{figure*}[t]
    \centering
    \includegraphics[scale=0.18]{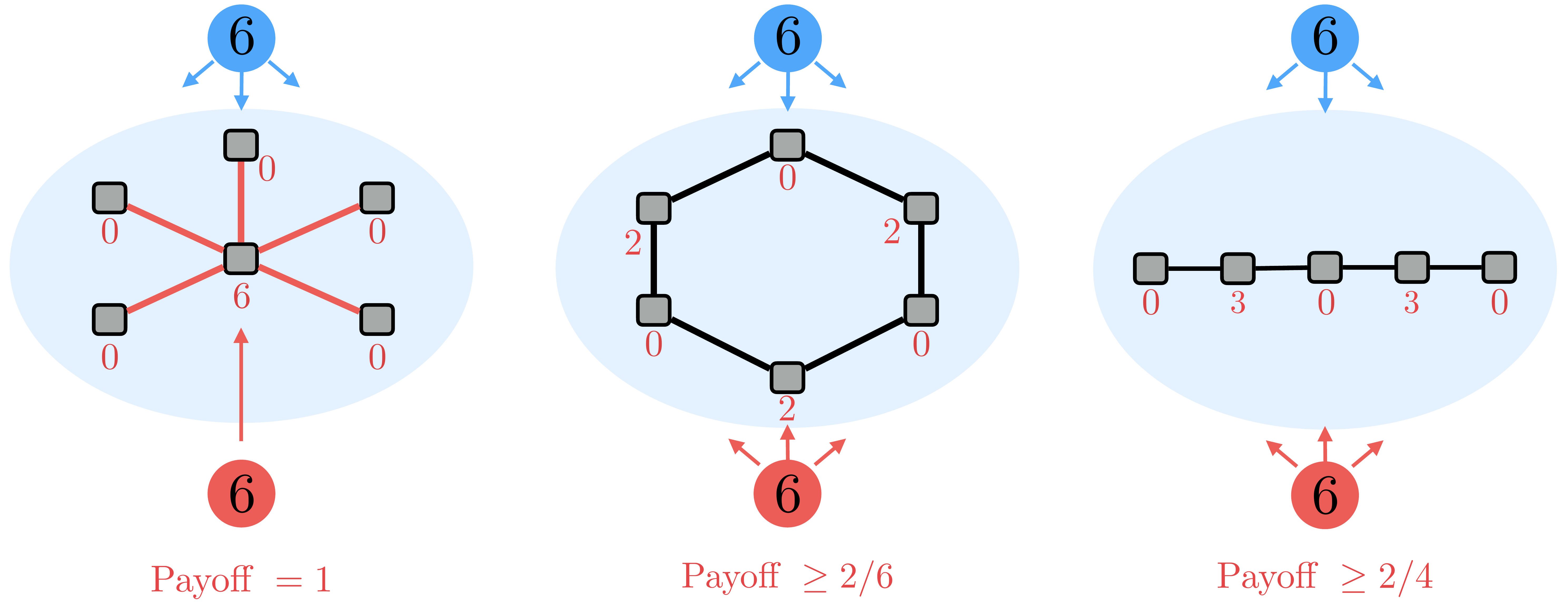}
    \caption{Three examples of a simplified setup where players are restricted to pure allocations. Here, we illustrate the max-min allocation strategy for the attacker (red) on three distinct networks -- that is, the highest payoff the attacker can guarantee regardless of the defender's (blue) allocation. Here, the defender and attacker each have 6 resource units, and we assume ties are awarded to the attacker. (Left) The attacker can guarantee that it secures every edge in the star network simply by sending all resources to the center node. There is no strategy that the defender can use to secure a single edge. (Center) On the ring graph of six nodes, the attacker can only guarantee that it secures two out of six edges. (Right) In the line network of five nodes, the attacker can only guarantee that it secures two out of four edges. In these  examples, the topology of the graph impacts the performance guarantees for the players.}
    \label{fig:bipartite_examples}
\end{figure*}

We further highlight the dependence of network structure by considering scenarios where the defender can only use deterministic strategies, as in Figure \ref{fig:bipartite_examples}. Here, we identify bipartite and complete graphs as the two extreme structures determining the range of the defender's effectiveness -- on complete graphs, a positive payoff is ensured against the widest range of attacker budgets compared to arbitrary graphs. On bipartite graphs, a positive payoff is ensured for the smallest range of attacker budgets (Proposition \ref{prop:deterministicstrategies}). These results are corroborated through numerical simulations on various random Erd{\"o}s-R{\'e}nyi graphs. Moreover, the simulations highlight the importance for the defender to implement randomized strategies, as there is a significant degradation in performance compared to the lower bound of performance in Theorem \ref{thm:BoundsGeneral}.

% We find that the players' equilibrium strategies, i.e. how players randomize their resource allocations over the nodes, are largely dependent on the degree centralities of the nodes. The resulting equilibrium payoffs, however, do not depend on any network characteristics.

% We then highlight the importance for the defender to implement randomized strategies by evaluating scenarios where it can only use deterministic strategies. We find that its performance in these scenarios is heavily dependent on the network structure. Through numerical simulations on random Erd{\"o}s-R{\'e}nyi graphs, we observe that its performance improves on highly connected networks -- it is worst on tree graphs and best on complete graphs.

\subsection{Related work}

A wide array of recent literature studies resource allocation over networks using a variety of different formulations in order to study the security of networked systems \cite{Shahrivar_2014,Guan_2019}. The impact of cyber attacks on dynamic networked systems is a central area of study in control and cyber-physical systems \cite{bai2017data,pasqualetti2015control,Milosevic_2020}. Game-theoretic approaches have focused on the problem of how a network of agents will invest in costly security protection resources \cite{Hota_2016_TCNS,Abdallah_2020}. The work of \cite{Abdallah_2020} considers a behavioral attack graph model and determines pure-strategy equilibrium security investments among multiple defenders of a network. Another formulation considers resources that can dynamically be re-allocated via network links \cite{shishika2022dynamic,chen2022path}.

The Colonel Blotto game and its variants are emerging as a flexible framework for studying complex adversarial interactions. As previously discussed, the most well-known results here are for settings where each player's objective is to accumulate as much value as possible by securing individual valuable battlefields \cite{Kovenock_handbook_2012}. As such,  alternate player objectives are a central focus in the Colonel Blotto literature, where success depends on securing subsets of battlefields rather than securing individual battlefields \cite{Shubik_1981,Golman_2009,Kovenock_handbook_2012,Shahrivar_2014,Kovenock_2018,Guan_2019,aghajan2023extension}. In \cite{Kovenock_2018}, a defender has a weakest-link objective associated with securing multiple networks, while the attacker has a best-shot objective associated with only securing a single network. Network Colonel Blotto games are considered in \cite{Guan_2019}, where a defender's payoff is determined by whether it can preserve certain network characteristics such as its connectivity or average degree.  This work primarily showcases computational methods to find approximate solutions in a model with integer-restricted allocations. In \cite{Shahrivar_2014}, pure strategy equilibria are found in a network formation Blotto game, where players succeed if they secure connected components of a graph. Our work contributes to this literature by featuring analytical characterizations of equilibrium mixed strategies in Network General Lotto games.

% To study the security of networked systems, another generalized variation of General Lotto game in which the battlefield are connected thorough a network are considered in the literature. More specifically,  consider a network whose nodes are battlefields, and the two players compete  to secure the network nodes. Here, the objective of each player depends on the structure of the network. For example, in , the authors consider a network General lotto game in which defender objective is to secure a set of specific nodes that is characterized by the structure of the network. In this model, we have a combination of weakest-link and best-shot game.
% In another model, in , the authors studied the model that defender try to keep the network connected while the attacker tries to destroy the connectivity of the network. To do so, the attacker tries to win cut-set while the defender defends them.
% One challenge to find an equilibrium General Lotto game over networks is that winning a node can potentially have multiple purpose. That's why, in , in general case, it was assumed that the different cut-sets are disjoint, and for only some special cases, the disjoint assumption was relaxed. 

\smallskip \noindent \textbf{Notation:} We denote $[n]=\{1,\ldots,n\}$ as the set of the first $n$ natural numbers. For a subset $B \subseteq [n]$ and a vector $x \in \R^n$, we write $x_B = (x_d)_{d\in B} \in \R^{|B|}$. We denote the set of non-negative real numbers as $\R_+$. We will use bold lettering to denote random variables, i.e. $\bx \sim F$ is a random variable with realization $x$ over some distribution $F$.

\section{Problem Formulation}\label{sec:formulation}

Consider a graph $G=(V,E)$, where $V \triangleq [n]$ is the set of vertices (or nodes) and $E \subseteq V \times V$ is the set of edges. We will use terms ``network" and ``graph" interchangeably. In this paper, we consider networks containing at least one edge, i.e., $E\not=\varnothing$. In a {\it Network General Lotto} game over the graph $G$, there are two opposing players $\mcx$ and $\mcy$, each aiming to secure as many edges as possible by allocating resources to the nodes.  In order to secure an edge $\{i,j\} \in E$, player $\mcx$ is required to out-compete $\mcy$, i.e. send more resources, on both nodes $i,j \in V$. Player $\mcy$ secures the edge $\{i,j\}$ if $\mcx$ fails to secure it. Indeed, it is more difficult for $\mcx$, who we refer to as the defender, to secure edges than it is for $\mcy$, who we refer to as the attacker.

% \begin{figure}
%     \centering
%     \includegraphics[scale=0.23]{figs/network_fig.pdf}
%     \caption{The left diagram depicts a Network Lotto game $(X,Y,G)$. Players $\mcx$ and $\mcy$ allocate resources against one another over the nodes of a graph $G$. The objective for each player is to secure as many edges of the network as possible. To secure an edge, the defender (player $\mcx$) is required to win both nodes (endpoints) associated with the edge. On the other hand, the attacker (player $\mcy$) only needs to win at least one of them. The right diagram illustrates an example outcome in which player $\mcx$ secures three edges by winning five nodes, and player $\mcy$ secures four edges by winning two nodes.}
%     \label{fig:setup}
% \end{figure} 

Formally, an allocation for player $\mcx$ is a vector $x = (x_i)_{i\in [n]} \in \mathbb{R}_{+}^{n}$, and similarly a vector $y\in \mathbb{R}_{+}^{n}$ for player $\mcy$. Player $\mcx$ (player $\mcy$) has a limited resource budget $X$ (budget $Y$) to allocate in expectation. An admissible strategy for $\mcx$ is any ${n}$-variate  (cumulative) distribution function $F_\mcx: \mathbb{R}_{+}^{n} \rightarrow [0,1]$ that belongs to
\begin{equation}\label{eq:LC}
    \F(X) \triangleq \left\{F_\mcx :  \E_{\bx \sim F_\mcx}\left[ \sum_{i\in [n]} \bx_i \right] \leq X \right\}.
\end{equation}
In words, player $\mcx$ can implement any randomization of allocations as long as it does not exceed its budget in \emph{expectation}. The admissible strategies belonging to \eqref{eq:LC} are the defining feature of General Lotto games relative to Colonel Blotto games \cite{Myerson_1993,Hart_2008,Kovenock_2021}: in Blotto games, the distribution $F_\mcx$ must not have support over any allocation that exceeds the budget \cite{Roberson_2006,Kovenock_2021}.

If $x$ and $y$ are allocations for players $\mcx$ and $\mcy$, the payoff awarded to player $\mcx$ is 
\begin{align}\label{eqn:piX}
  \pi_\mcx(x,y;G)\triangleq \frac{1}{|E|}\sum_{\{i,j\} \in E}1_{\{x_i\geq y_i,x_j\geq y_j\}},
\end{align}
where 
\begin{align*}
    1_{\{x_i\geq y_i,x_j\geq y_j\}}\triangleq\begin{cases}
        1,& \text{if }x_i\geq y_i,x_j\geq y_j\\
        0,& \text{otherwise}
    \end{cases}.
\end{align*}
In other words, player $\mcx$ secures the edge $\{i,j\}$, if he wins {\it both} nodes $i,j$. Conversely, player $\mcy$ secures the edge $\{i,j\}$ if he wins at least one of the nodes $i,j$. Therefore, 
the payoff awarded to player $\mcy$ is $\pi_\mcy(x,y)=1-\pi_\mcx(x,y)$, i.e. it is a constant-sum game. Given a strategy profile $(F_\mcx,F_\mcy) \in \F(X)\times \F(Y)$, the \emph{expected payoffs} to each player are denoted as 
\begin{equation}
    \begin{aligned}
        \pi_\mcx(F_\mcx,F_\mcy;G) &= \E_{\bx \sim F_\mcx,\by\sim F_\mcy}[\pi_\mcx(\bx,\by;G)], \\ 
        \pi_\mcy(F_\mcx,F_\mcy;G) &= \E_{\bx \sim F_\mcy,\by\sim F_\mcy}[\pi_\mcy(\bx,\by;G)].
    \end{aligned}
\end{equation}
We will denote a Network General Lotto game over a graph $G$ with budgets $X,Y$ as the triple $(X,Y,G)$.

\begin{definition}
    An \it{equilibrium} of $(X,Y,G)$ is a strategy profile $(F_\mcx^*,F_\mcy^*) \in \F(X)\times \F(Y)$ that satisfies
    \begin{align}\label{eqn:equilibriumdefinition}
        \pi_\mcx(F_\mcx,F_\mcy^*;G) \leq \pi_\mcx(F_\mcx^*,F_\mcy^*;G) \leq \pi_\mcx(F_\mcx^*,F_\mcy;G)
        % \pi_\mcx(F_\mcx^*,F_\mcy^*;G) &\geq \pi_\mcx(F_\mcx,F_\mcy^*;G),\cr
        % \pi_\mcy(F_\mcx^*,F_\mcy^*;G) &\geq \pi_\mcy(F_\mcx^*,F_\mcy;G)
    \end{align}
for all $F_\mcx \in \F(X)$ and $F_\mcy \in \F(Y)$.
\end{definition}
 
We will often omit the dependence of $\pi_\mcx$ on the graph $G$ when the context is clear.
% It is challenging to derive equilibrium strategies and payoffs for $\NL(X,Y,G)$ in the most general cases, i.e. for arbitrary graphs $G$. In this paper, we completely characterize equilibria for the class of bipartite graphs.

\section{Bipartite networks}\label{sec:bipartite}

In this section, we will restrict attention to the class of bipartite graphs. The definition of a bipartite graph is given below:

\begin{definition}
    A graph $G = (V,E)$ is \emph{bipartite} if there are two disjoint subsets of nodes $B_1,B_2$ such that $B_1 \cup B_2 = V$, and no edges exist between nodes in the same subset.
\end{definition}
Equivalently, a bipartite graph is a graph that does not contain any odd-length cycles. Recall that the example networks from Figure \ref{fig:bipartite_examples}, i.e. star, ring (even number of nodes), and line, were all bipartite. In the simplified setting from the examples where only deterministic integer allocations were permitted, the players' performance guarantees heavily depended on the network's structure.

\subsection{Equilibrium characterizations}

We present our main result below, which is an equilibrium characterization of the Network Lotto game for any bipartite network.

\begin{figure*}[t]
    \centering
    \includegraphics[scale=0.22]{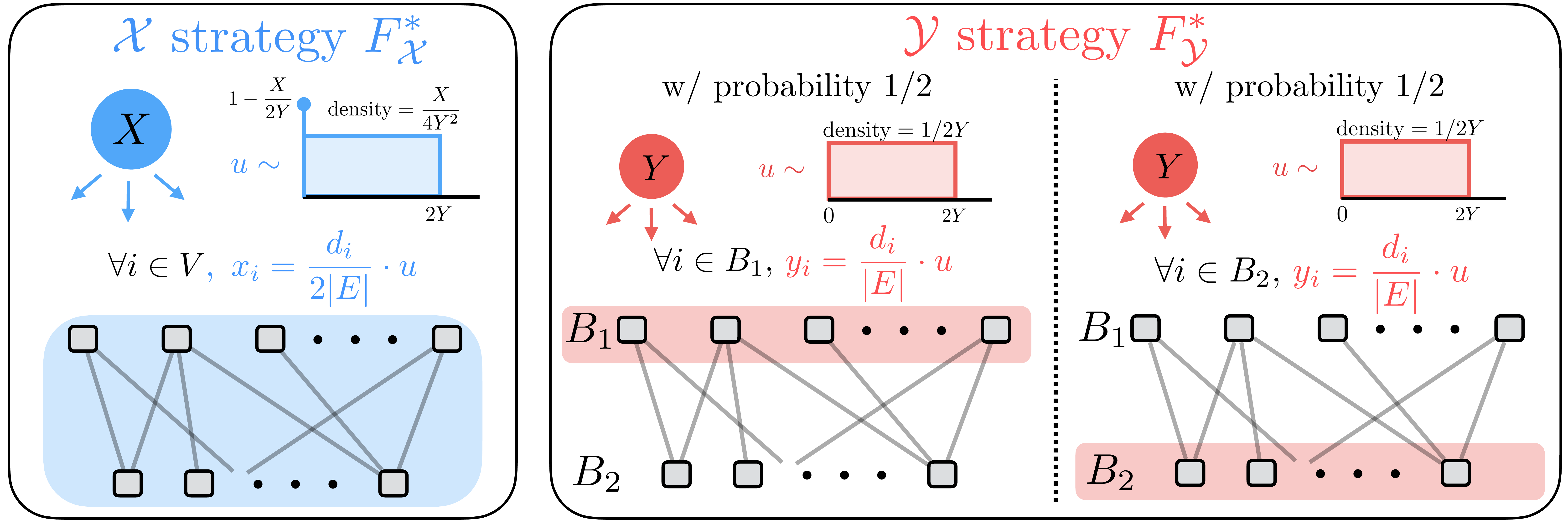}
    \caption{Illustration of equilibrium strategies on bipartite networks (Theorem \ref{thm:NL2}) in the regime $X < 2Y$. The case $X \geq 2Y$ is similar. (Left) The defender's equilibrium strategy is to draw a sample $u$ from the density function shown. Using a total of $u$ resources, it distributes to each node in the network proportionally to their degree centralities, i.e. $x_i = \frac{d_i}{2|E|}u$. Observe that with probability $1-\frac{X}{2Y}$, no resources are allocated at all. (Right) The attacker's equilibrium strategy is to first choose one of the node partitions $B_k$, $k \in \{1,2\}$ with probability 1/2. It then draws a sample $u$ from the uniform density function shown. Using a total of $u$ resources, it distributes to each node $i$ in $B_k$ proportionally to their degree centralities, i.e. $y_i= \frac{d_i}{|E|}u$. For each node $i \notin B_k$, no resources are allocated. The payoffs that result from this strategy profile is identical for every bipartite network.}
    \label{fig:bipartite_strategies}
\end{figure*}

\begin{theorem}\label{thm:NL2}
   Consider a Network General Lotto game $(X,Y,G)$, where $G$ is bipartite. The equilibrium payoff to $\mcx$ is given by
    \begin{equation}\label{eq:piXstar}
        \gamma(X,Y) \triangleq 
        \begin{cases}
            1 - \frac{Y}{X}, &\text{if } X \geq 2Y \\
            \frac{X}{4Y}, &\text{if } X < 2Y \\
        \end{cases},
    \end{equation}
    and the payoff to $\mcy$ is $1 - \gamma(X,Y)$. An equilibrium strategy profile $(F_\mcx^*,F_\mcy^*)$ is given as follows. Let $B_1,B_2$ be the bipartite partition of $G$.
    
    \noindent If $X < 2Y$: 
    \begin{equation}\label{eq:thmstrat_weaker}
        \begin{aligned}
            F_\mcx^*(x)&=1-\frac{X}{2Y}+\frac{X|E|}{4Y^2}\min\left\{\left\{ \frac{x_i}{d_i} \right\}_{i\in V},\frac{2Y}{|E|} \right\} \\
            F_\mcy^*(y)&=\frac{|E|}{4Y}\sum_{k=1,2} \min\left\{ \left\{\frac{y_i}{d_i}\right\}_{i\in B_k}, \frac{2Y}{|E|} \right\}
        \end{aligned}
    \end{equation}
    
    \noindent If $X\geq 2Y$: 
    \begin{equation}\label{eq:thmstrat_stronger}
        \begin{aligned}
            F_\mcx^*(x)&=\frac{|E|}{X}\min \left\{ \left\{\frac{x_i}{d_i}\right\}_{i\in V},\frac{X}{|E|} \right\} \\
            F_\mcy^*(y)&=1-\frac{2Y}{X}+\frac{Y|E|}{X^2}\sum_{k=1,2}\min\left\{\left\{\frac{y_i}{d_i}\right\}_{i\in B_k},\frac{X}{|E|}\right\}
        \end{aligned}
    \end{equation}
\end{theorem}
The equilibrium payoff \eqref{eq:piXstar} is \emph{identical} for any bipartite graph $G$, and are hence \emph{independent} of any other graph characteristics. It only depends on the relative budgets $X$ and $Y$. The equilibrium strategies $(F_\mcx^*,F_\mcy^*)$, however, are strongly linked to the particular structure of the bipartite graph. We elaborate on the equilibrium strategies specified by \eqref{eq:thmstrat_weaker} and \eqref{eq:thmstrat_stronger} below.

\subsection{Interpretation of equilibrium strategies}

For the case $X<2Y$ \eqref{eq:thmstrat_weaker}, the defender's strategy $F_\mcx^* \in \F(X)$  randomizes over allocations as follows. With probability $1-\frac{X}{2Y}$, no resources are allocated at all. With probability $\frac{X}{2Y}$, a single sample $u\sim \text{Unif}[0,4Y]$ is drawn, and the allocation is determined as $x_i = \frac{d_i}{2|E|}\cdot u$ for each node $i \in V$. In other words, $\mcx$ allocates a total of $u$ resources among the nodes proportionally to their degree centralities. The allocations to each node $x_i$ are random, but correlated through the uniform sample $u$. One can verify that this strategy is budget-feasible in expectation. An illustration is depicted in Figure \ref{fig:bipartite_strategies} (left).

The attacker's strategy $F_\mcy^* \in \F(Y)$ randomizes over allocations as follows. With probability 1/2, it allocates resources only to nodes in the partition $B_k$, $k \in \{1,2\}$, in the following manner. A single sample $u \sim \text{Unif}[0,2Y]$ is drawn. The allocation to each node $i \in B_k$ is determined as $y_i = \frac{d_i}{|E|}\cdot u$ and to each node $i \in B_{-k}$ as $y_i = 0$. An illustration is depicted in Figure \ref{fig:bipartite_strategies} (right).

A similar interpretation applies for the case $X \geq 2Y$. The difference here is that $\mcx$ is the ``stronger" player, and will not give up (allocate nothing) with a non-zero probability. As the ``weaker" player, $\mcy$ assigns a non-zero probability to give up. 

Revisiting the examples from Figure \ref{fig:bipartite_examples}, we can begin to reason why the initial intuition would not apply to the full Network Lotto game. Considering the star network, even if the attacker has more resources than the defender, and allocates only to the center node, the defender is able to counter this strategy because the admissible strategy space $\F(X)$ enables it to randomize over allocations that exceed the attack on the center. Thus, the attacker itself needs to randomize its allocation to other nodes in the network as well.

\section{Beyond Bipartite Graphs}\label{sec:beyond}

In this section, we investigate the players' performance on graphs that are not bipartite. While we do not provide equilibrium characterizations for arbitrary graphs $G$, we seek to provide bounds on the defender's optimal security payoff:
\begin{equation}\label{eq:security_value_X}
    S_\mcx^*(X,Y,G) \triangleq \max_{F_\mcx \in \F(X)}\min _{F_\mcy \in \F(Y)}\pi_\mcx(F_\mcx,F_\mcy;G)
\end{equation}

In the result below, we establish lower and upper bounds on player $\mcx$'s optimal security payoff for arbitrary graphs.
\begin{theorem}\label{thm:BoundsGeneral}
    Consider a Network General Lotto game $(X,Y,G)$, where $G=(V,E)$ is any graph with $n = |V|\geq 2$ nodes. Then
    \begin{equation}\label{eq:maxmin_bound}
        \gamma(X,Y) \leq S_\mcx^*(X,Y,G)\leq \gamma_n(X,Y)
    \end{equation}
    where $\gamma(X,Y)$ is defined in \eqref{eq:piXstar}, and 
    \begin{equation}\label{eq:general_UB}
        \gamma_n(X,Y) \triangleq
        \begin{cases}
            1 - \frac{n}{2(n-1)}\frac{Y}{X}, &\text{if } X \geq \frac{n}{n-1}Y \\
            \frac{n-1}{2n}\frac{X}{Y}, &\text{if } X < \frac{n}{n-1}Y
        \end{cases}
    \end{equation}
    for $n = 2,3,\ldots$.
\end{theorem}

\begin{figure}[t]
    \centering
    \includegraphics[scale=0.23]{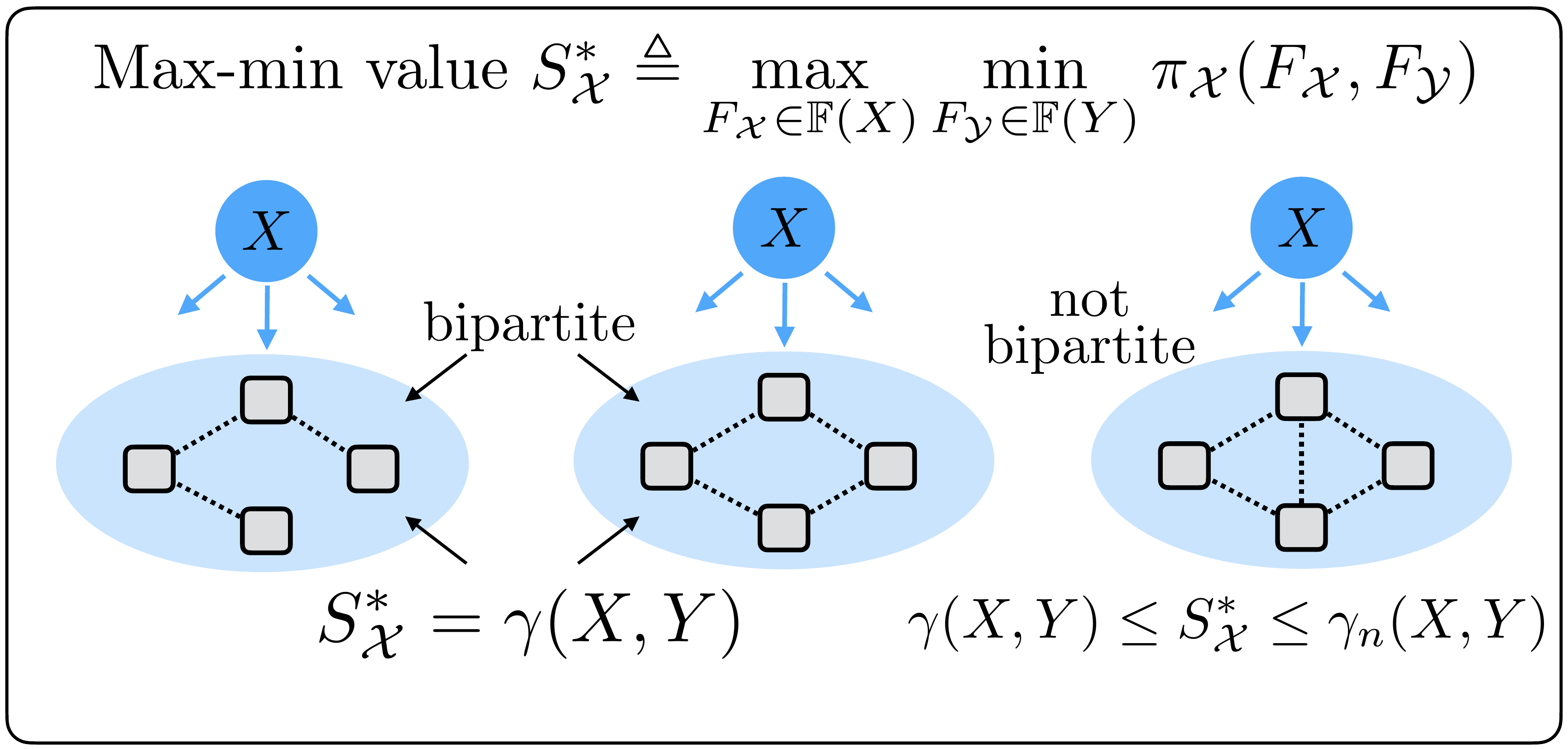}
    \caption{Theorem \ref{thm:NL2} states $\gamma(X,Y)$ \eqref{eq:piXstar} is the equilibrium payoff to the defender $\mcx$ on any bipartite network (left and center networks). The right network is not bipartite -- Theorem \ref{thm:BoundsGeneral} states upper and lower bounds on the max-min value for $\mcx$.}
    \label{fig:thm_fig}
\end{figure}

There are several interesting things to note in Theorem \ref{thm:BoundsGeneral}. Player $\mcx$ can find a strategy $F_\mcx \in \F(X)$ that guarantees itself a payoff of at least $\gamma(X,Y)$, regardless of the graph $G$ and $\mcy$'s strategy. The fact that $\mcx$ can ensure a payoff of at least $\gamma(X,Y)$ on any network suggests that bipartite graphs are the ``most difficult" to defend. As such, Theorem \ref{thm:BoundsGeneral} implies that the optimal performance guarantee cannot decrease by adding any amount of additional edges to an existing bipartite network. Thus, only the defender can benefit from adding edges to the network, e.g. by forming cliques or odd-length cycles. An illustration summarizing Theorems \ref{thm:NL2} and \ref{thm:BoundsGeneral} is provided in Figure \ref{fig:thm_fig}.

\begin{figure}
    \centering
    \includegraphics[scale=0.2]{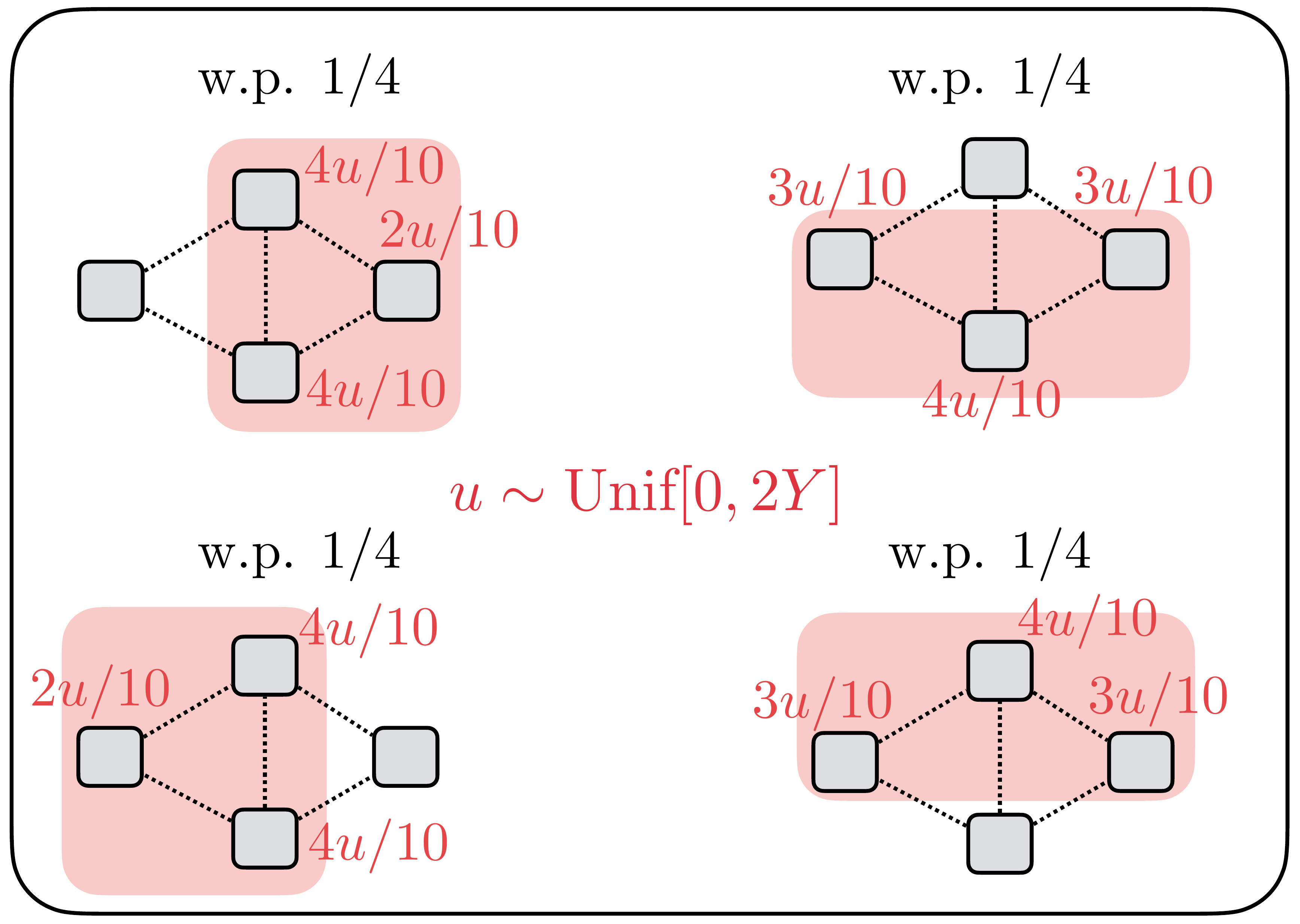}
    \caption{Illustration of the attacker's strategy on arbitrary graphs that places the upper bound $\gamma_n$ on the defender's attainable payoff (Theorem \ref{thm:BoundsGeneral}). Resources are allocated proportionally according to \eqref{eq:UB_weights1}. Here, $u \sim \text{Unif}[0,2Y]$ is the randomization on total resources.}
    \label{fig:UB_strategy}
\end{figure}

The upper bound of \eqref{eq:maxmin_bound} indicates that $\mcy$ can find a strategy $F_\mcy \in \F(Y)$ that ensures $\mcx$ cannot obtain a payoff that exceeds $\gamma_n(X,Y)$. We informally describe such a strategy now and in Figure \ref{fig:UB_strategy}. A detailed analysis is provided in Section \ref{sec:bounds_proof}. For any graph $G=(V,E)$, consider the collection of $n$ vertex covers $V_i = V\backslash\{i\}$. With probability $1/n$, the strategy $F_\mcy$ selects the vertex cover $V_k$, $k\in V$, and allocates a total of $u \sim \text{Unif}[0,2Y]$ resources among nodes $i \in V_k$ proportionally according to
\begin{equation}\label{eq:UB_weights1}
    y_i = \frac{u}{2|E|}\cdot
    \begin{cases}
        d_i, &\text{if } i \notin \mcal{N}_k \\
        d_i+1, &\text{if } i \in \mcal{N}_k
    \end{cases}
\end{equation}
The first entry above indicates that the share of resources sent to nodes not connected to $k$ is proportional to their degree centralities. The second entry above places slightly more weight to nodes that are direct neighbors of $k$. The motivation to add extra weight to these nodes is because each one uniquely covers the edges $\{i,k\}$, $i \in \mcal{N}_k$. There is more redundancy for any other edge $\{i,j\}$ in the graph, since at least two nodes $i,j \notin \mcal{N}_k$ are able to cover it.

\section{Analysis and proofs}\label{sec:proof}
% !TEX root = main.tex

In this section, we highlight analytical techniques and provide the proofs for our main results. 

\subsection{Correlated allocation strategies}
Before proceeding with the proofs, we first define a class of randomized allocation strategies which we term \emph{correlated allocation strategies}. This class of strategies serves as the basis for our analysis, and generalizes the randomized allocation strategies that were detailed in Sections \ref{sec:bipartite} and \ref{sec:beyond}.

\begin{definition}
    Consider a player $\mcz \in \{\mcx,\mcy\}$ with resource budget $Z$, and a finite collection of subsets of nodes $\mcd \subseteq 2^V$. A \emph{correlated allocation strategy} is specified by a cumulative distribution function $F_\mcz \in \F(Z)$ for $z \in \R_+^{|V|}$ of the form
    \begin{equation}\label{eq:correlated_strat}
        \begin{aligned}
            F_\mcz(z) &= 1 - \delta + \frac{\delta^2}{2Z}\sum_{D \in \mcd} p_D  \min\left\{\left\{\frac{z_i}{w_{D,i}} \right\}_{i \in D}, \frac{2Z}{\delta}\right\},
        \end{aligned}
    \end{equation}
    for some $\delta \in [0,1]$, probabilities $\{p_D\}_{D\in\mcd}$ s.t. $\sum_{D\in\mcd} p_D = 1$, and positive weights $\{w_{D,i}\}_{i\in D}$ for each $D \in \mcd$ s.t. $ \sum_{i\in D} w_{D,i}=1$.
\end{definition}
A correlated allocation strategy can more intuitively be described as follows. To generate a sample allocation $\bz \sim F_\mcz$, player $\mcz$ first randomly selects a subset of nodes $D$ from the collection $\mcd$ according to the probability vector $\{p_D\}_{D\in\mcd}$. Player $\mcz$ then allocates resources only to nodes $i\in D$ in the following fashion. With probability $1-\delta$, no resources are allocated at all. With probability $\delta$, a single sample from the uniform distribution on the interval $\left[0,\frac{2Z}{\delta}\right]$ is taken, i.e. $\mathbf{u} \sim \text{Unif}\left[0,\frac{2Z}{\delta}\right]$. Then, the resources are allocated according to $\bz_i = w_{D,i}\cdot \mathbf{u}$ for $i \in D$ and $\bz_i = 0$ for $i\notin D$. It follows that the strategy \eqref{eq:correlated_strat} is budget-feasible, i.e. $\E_{\bz \sim F_\mcz}\left[\sum_{i=1}^n \bz_i\right] = Z$.

Thus, the random variable $\mathbf{u}$ serves as a correlating device on the player's allocations to nodes in $D$, where the amounts are proportional to the weights $w_{D,i}$. We identify the following structural property regarding best responses against correlated allocation strategies.
\begin{lemma}\label{lem:support_exceed}
    Consider a correlated allocation strategy $F_\mcx \in \F(X)$ for player $\mcx$ of the form \eqref{eq:correlated_strat}. Then for any $F_\mcy \in \F(Y)$, there exists a $F_\mcy' \in \F(Y)$ such that
    \begin{itemize}
        \item $\pi_\mcy(F_\mcy',F_\mcx) \geq \pi_\mcy(F_\mcy,F_\mcx)$, and
        \item for all $i\in V$ s.t. $i \in D$ for some $D \in \mcd$, we have 
        \begin{align}\label{lem:InsideEqu}
            \mbb{P}_{\by\sim F_\mcy'}\left(\by_i > \frac{2X\max_{D\in\mcd} \{w_{D,i}\}}{\delta}  \right) = 0.
        \end{align}
    \end{itemize}
    Here, we assume $w_{D,i} = 0$ if $i \notin D$. An identical result holds with player indices reversed.
\end{lemma}
% \pf If $F_\mcy$ satisfies \eqref{lem:InsideEqu}, then there is nothing to prove. Therefore, assume that $F_\mcy$ does not satisfy \eqref{lem:InsideEqu} for some $i\in V$. Let 

In words, a player prefers not to randomize over allocations outside of the support of the other player's correlated allocation strategy. This is because for some distribution $F_\mcy \in \F(Y)$ such that $\mbb{P}_{\by\sim F_\mcy}\left(\by_i > \frac{2X\max_{D\in\mcd} \{w_{D,i}\}}{\delta}  \right) > 0$, we can use another distribution $F'_\mcy$ such that 
\begin{align*}
            F_\mcy'\left(y  \right) = \begin{cases}
                F(y),& y<\frac{2X\max_{D\in\mcd} \{w_{D,i}\}}{\delta}\\
                1,& \text{otherwise}
            \end{cases},
        \end{align*}
        without losing any payoff.
% We start with the proof of Theorem \ref{thm:NL2}. 

\subsection{Equilibrium characterizations on bipartite graphs}

Consider a bipartite graph $G=(V,E)$ with partition $B_1,B_2$. Let $d_i$ be the degree of node $i\in V$. Recall the profile $(F_\mcx^*,F_\mcy^*)$ of correlated allocation strategies defined in \eqref{eq:thmstrat_weaker} and \eqref{eq:thmstrat_stronger}. To reiterate,

\noindent $\bullet$ If $X< 2Y$: $\mcx$ uses the correlated allocation strategy with $\mcd = \{V\}$, $\delta = \frac{X}{2Y}$, and $w_{V,i} = \frac{d_i}{2|E|}$ for $i\in V$. From \eqref{eq:correlated_strat}, we recover
\begin{align}\label{eqn:thmFX2}
    F_\mcx^*(x)=1-\frac{X}{2Y}+\frac{X|E|}{4Y^2}\min\left\{\left\{ \frac{x_i}{d_i} \right\}_{i\in V},\frac{2Y}{|E|} \right\}.
\end{align}
Player $\mcy$ uses the correlated allocation strategy with $\mcd = \{B_1,B_2\}$, $p_{B_k} = 1/2$ for $k\in\{1,2\}$, $\delta = 1$, and $w_{B_k,i} = \frac{d_i}{|E|}$ for $k\in\{1,2\}$ and $i\in B_k$. From \eqref{eq:correlated_strat}, we recover

\begin{align}\label{eqn:thmFY2}
    F_\mcy^*(y)=\frac{|E|}{4Y}\sum_{k=1,2} \min\left\{ \left\{\frac{y_i}{d_i}\right\}_{i\in B_k}, \frac{2Y}{|E|} \right\}.
\end{align}

\noindent $\bullet$ If $X\geq 2Y$: $\mcx$ uses the correlated allocation strategy with $\mcd = \{V\}$, $\delta = 1$, and $w_{V,i} = \frac{d_i}{2|E|}$ for $i\in V$. From \eqref{eq:correlated_strat},
\begin{align}\label{eqn:thmFX1}
    F_\mcx^*(x)=\frac{|E|}{X}\min \left\{ \left\{\frac{x_i}{d_i}\right\}_{i\in V},\frac{X}{|E|} \right\}.
\end{align}
Player $\mcy$ uses the correlated allocation strategy with ${\mcd = \{B_1,B_2\}}$, $p_{B_k} = 1/2$ for $k\in\{1,2\}$, $\delta = \frac{2Y}{X}$, and $w_{B_k,i} = \frac{d_i}{|E|}$ for $k\in\{1,2\}$ and $i\in B_k$. From \eqref{eq:correlated_strat},
% for $x\in \prod_{i=1}^n\left[0,\frac{d_i X}{|E|}\right]$
\begin{align}\label{eqn:thmFY1}
    F_\mcy^*(y)=1-\frac{2Y}{X}+\frac{Y|E|}{X^2}\sum_{k=1,2}\min\left\{\left\{\frac{y_i}{d_i}\right\}_{i\in B_k},\frac{X}{|E|}\right\}.
\end{align}
We are now ready to prove Theorem \ref{thm:NL2}.

{\it Proof of Theorem \ref{thm:NL2}:}
To prove the theorem, we proceed with the following steps:
\begin{enumerate}
    \item We prove for every $F_\mcx\in\F(X)$, $\pi_\mcx(F_\mcx,F^*_\mcy)\leq \pi^*_\mcx(X,Y)$ for some $\pi^*_\mcx(X,Y)\in\R_+$ which is the payoff of player $\mcx$.
    \item Then, we show $\pi_\mcx(F^*_\mcx,F^*_\mcy)= \pi^*_\mcx(X,Y)$.  
    \item Finally, we prove  $\pi_\mcy(F^*_\mcx,F_\mcy)\leq 1-\pi^*_\mcx(X,Y)$ for every $F_\mcy\in\F(Y)$.
\end{enumerate}
Since $\pi_\mcy(F^*_\mcx,F^*_\mcy)=1-\pi_\mcx(F^*_\mcx,F^*_\mcy)= 1-\pi^*_\mcx(X,Y)$, we also have $\pi_\mcy(F_\mcx,F_\mcy)= 1- \pi^*_\mcx(X,Y)$.  Therefore,   from the definition of equilibrium \eqref{eqn:equilibriumdefinition}, proving these steps shows that $(F^*_\mcx,F^*_\mcy)$ is an equilibrium with equilibrium payoff $\pi^*_\mcx(X,Y)$ for player $\mcx$ and payoff $1 - \pi^*_\mcx(X,Y)$ for player $\mcy$.

We start the proof for the case $X\geq 2Y$. For any distribution $F_\mcx\in\F(X)$ for player $\mcx$, we have
\begin{align*}
   \pi_\mcx(F_\mcx,F^*_\mcy)&=\E_{F_\mcx,F^*_\mcy}[\pi_\mcx(\bx,\by)]\cr
    &=\E_{F_\mcx}\left[\E_{F^*_\mcy}[\pi_\mcx(\bx,\by)\mid \bx]\right],
        \end{align*}
        which follows from the law of total expectation. From above equality and the payoff function for player $\mcx$ \eqref{eqn:piX}, we have
    \begin{align*}
    \pi_\mcx(&F_\mcx,F^*_\mcy)\cr
    &=\E_{F_\mcx}\left[\E_{F^*_\mcy}\left[\frac{1}{|E|}\sum_{\{i,j\} \in E} 1_{\{\bx_i\geq \by_i,\bx_j\geq \by_j\}}\bigg{|} \bx\right]\right]\cr
    &=\E_{F_\mcx}\left[\frac{1}{|E|}\sum_{\{i,j\} \in E} \E_{F^*_\mcy}\left[1_{\{\bx_i\geq \by_i,\bx_j\geq \by_j\}}\bigg{|} \bx\right]\right]\cr
    &=\E_{F_\mcx}\left[\frac{1}{|E|}\sum_{\{i,j\} \in E} \E_{F^*_{\mcy,\{i,j\}}}\left[1_{\{\bx_i\geq \by_i,\bx_j\geq \by_j\}}\bigg{|} \bx\right]\right],
    \end{align*}
    which follows from the fact that $1_{\{\bx_i\geq \by_i,\bx_j\geq \by_j\}}$ does not depend on $\bx_{[n]\setminus\{i,j\}},\by_{[n]\setminus\{i,j\}}$. Here, we denote $F_{\mcz,e}$, where $\mcz \in \{\mcx,\mcy\}$ and  $e=\{i,j\}\in E$, as the (two-dimensional) marginal distribution of $F_\mcz$ over $\bx_i,\bx_j$. Therefore, from above equality, we have
    \begin{align*}
    \pi_\mcx(F_\mcx,F^*_\mcy)
    &=\E_{F_\mcx}\left[\frac{1}{|E|}\sum_{\{i,j\} \in E}\PP(\bx_i\geq \by_i,\bx_j\geq \by_j|\bx)\right]\cr
    &=\E_{F_\mcx}\left[\frac{1}{|E|}\sum_{\{i,j\} \in E} F^*_{\mcy,\{i,j\}}(\bx_{\{i,j\}})\right].
    \end{align*}
    From \eqref{eqn:thmFY1}, the bi-variate marginal distribution w.r.t allocation to nodes $i$ and $j$ is
    \begin{align*}
        F&^*_{\mcy,\{i,j\}}(x_{\{i,j\}})\cr
        &=F^*_\mcy(x)\big{|}_{x_k=\infty, k\in[n]\setminus{\{i,j\}}}\cr
        &=1-\frac{2Y}{X}+\frac{Y|E|}{X^2}\left(\min\left\{\frac{x_i}{d_i},\frac{X}{|E|}\right\}+\min\left\{\frac{x_j}{d_j},\frac{X}{|E|}\right\}\right)\cr
        &=1-\frac{2Y}{X}+\frac{Y|E|}{X^2}\left(\frac{x_i}{d_i}+\frac{x_j}{d_j}\right).
    \end{align*}
    The last equality follows from Lemma \ref{lem:support_exceed}. We then have
    \begin{align*}
    \pi_\mcx&(F_\mcx,F^*_\mcy)\cr
    &=\E_{F_\mcx}\left[\frac{1}{|E|}\sum_{ \{i,j\} \in E} \left(1-\frac{2Y}{X}+\frac{Y|E|}{X^2}\left(\frac{\bx_i}{d_i}+\frac{\bx_j}{d_j}\right)\right)\right]\cr
    &=\E_{F_\mcx}\left[1-\frac{2Y}{X}+\frac{Y}{X^2}\sum_{\{i,j\} \in E} \left(\frac{\bx_i}{d_i}+\frac{\bx_j}{d_j}\right)\right].
        \end{align*}
        Since 
        \begin{align}\label{eqn:xidi}
            \sum_{\{i,j\} \in E} \left(\frac{x_i}{d_i}+\frac{x_j}{d_j}\right)=\sum_{i\in [n]} d_i\left(\frac{x_i}{d_i}\right)=\sum_{i\in [n]}x_i,
        \end{align}
        the above equality implies 
    \begin{align*}
    \pi_\mcx(F_\mcx,F^*_\mcy)
    &=\E_{F_\mcx}\left[1-\frac{2Y}{X}+\frac{Y}{X^2} \sum_{i=1}^n\bx_i\right]\cr
    &\stackrel{(a)}{\leq}1-\frac{2Y}{X}+\frac{Y}{X} \cr
    &= 1-\frac{Y}{X},
\end{align*}
where $(a)$ follows from $\E_{F_\mcx}\left[\sum_{i=1}^n\bx_i\right]\leq X$. Also, note that since $\E_{F^*_\mcx}\left[\sum_{i=1}^n\bx_i\right]= X$, we have $\pi_\mcx(F^*_\mcx,F^*_\mcy)=1-\frac{Y}{X}$.  

Similarly, for any distribution $F_\mcy\in\F(Y)$ for player $\mcy$, we have
\begin{align}\label{eqn:PPbxigeqbyibxjgeqbyjby}
   \pi_\mcy(&F^*_\mcx,F_\mcy)\cr
   &=1-\E_{F^*_\mcx,F_\mcy}[\pi_\mcx(\bx,\by)]\cr
    &=1-\E_{F_\mcy}\left[\E_{F^*_\mcx}[\pi_\mcx(\bx,\by)\mid \bx]\right]\cr
    % &=1-\E_{F'_\mcy}\left[\E_{F_\mcy}\left[\sum_{e \in E}w_e \prod_{i\in e}1_{\{x_i\geq y_i\}}\bigg{|} \bx\right]\right]\cr
    &=1-\E_{F_\mcy}\left[\frac{1}{|E|}\sum_{\{i,j\} \in E} \E_{F^*_{\mcx,\{i,j\}}}\left[1_{\{\bx_i\geq \by_i,\bx_j\geq \by_j\}}\bigg{|} \by\right]\right]\cr
    &=1-\E_{F_\mcy}\left[\frac{1}{|E|}\sum_{\{i,j\} \in E}\PP(\bx_i\geq \by_i,\bx_j\geq \by_j|\by)\right].
        \end{align}
        Moreover, we have
        \begin{align*}
          \PP(\bx_i\geq \by_i,\bx_j\geq \by_j|\by)&=1+F^*_{\mcx,\{i,j\}}(\by_i,\by_j)\cr
          &\quad-F^*_{\mcx,\{i\}}(\by_i)-F^*_{\mcx,\{j\}}(\by_j)\cr
          &\stackrel{(a)}{=}1+\frac{|E|}{X}\min\left\{\frac{\by_i}{d_i},\frac{\by_j}{d_j}\right\}\cr
          &\quad-\frac{|E|}{X}\frac{\by_i}{d_i}-\frac{|E|}{X}\frac{\by_j}{d_j}\cr
          &=1-\frac{|E|}{X}\max\left\{\frac{\by_i}{d_i},\frac{\by_j}{d_j}\right\},
        \end{align*}
        where $(a)$ follows from \eqref{eqn:thmFX1}.
        Plugging above equality into \eqref{eqn:PPbxigeqbyibxjgeqbyjby} gives
    \begin{align*}
    \pi_\mcy(F^*_\mcx,F_\mcy)&=\E_{F_\mcy}\left[\frac{1}{X}\sum_{\{i,j\} \in E} \max\left\{\frac{\by_i}{d_i},\frac{\by_j}{d_j}\right\}\right]\cr
    &\leq\E_{F_\mcy}\left[\frac{1}{X}\sum_{\{i,j\} \in E}\left(\frac{\by_i}{d_i}+\frac{\by_j}{d_j}\right)\right]\cr
    &\stackrel{(a)}{=}\E_{F_\mcy}\left[\frac{1}{X}\sum_{i=1}^n {\by_i}\right]\cr
    &\stackrel{(b)}{\leq}\frac{Y}{X},
\end{align*}
where $(a)$ follows from the same argument as \eqref{eqn:xidi}, and $(b)$ follows from $\E_{F_\mcy}\left[\sum_{i=1}^n\by_i\right]\leq Y$.

The proof for the case $X<2Y$ follows the similar idea to  the case $X\geq 2Y$. For completeness, we provide the main lines of the proof for the case $X<2Y$. For any $F_\mcx\in\F(X)$, we have
\begin{align*}
   \pi_\mcx(&F_\mcx,F^*_\mcy)\cr
    &=\E_{F_\mcx}\left[\E_{F^*_\mcy}\left[\frac{1}{|E|}\sum_{\{i,j\} \in E} 1_{\{\bx_i\geq \by_i,\bx_j\geq \by_j\}}\bigg{|} \bx\right]\right]\cr
    &=\E_{F_\mcx}\left[\frac{1}{|E|}\sum_{\{i,j\} \in E} \E_{F^*_{\mcy,\{i,j\}}}\left[1_{\{\bx_i\geq \by_i,\bx_j\geq \by_j\}}\bigg{|} \bx\right]\right]\cr
    &=\E_{F_\mcx}\left[\frac{1}{|E|}\sum_{\{i,j\} \in E} F^*_{\mcy,\{i,j\}}(\bx_{\{i,j\}})\right]\cr
    &=\E_{F'_\mcx}\left[\frac{1}{|E|}\sum_{ \{i,j\} \in E} \frac{|E|}{4Y}\left(\frac{\bx_i}{d_i}+\frac{\bx_j}{d_j}\right)\right]\cr
    &=\E_{F'_\mcx}\left[\frac{1}{4Y} \sum_{i=1}^n\bx_i\right]
    \leq \frac{X}{4Y}.
\end{align*}

For any $F_\mcy\in\F(Y)$, we have
\begin{align*}
   \pi_\mcy(&F^*_\mcx,F_\mcy)\cr
    &=1-\E_{F_\mcy}\left[\frac{1}{|E|}\sum_{\{i,j\} \in E} \E_{F^*_{\mcx,\{i,j\}}}\left[1_{\{\bx_i\geq \by_i,\bx_j\geq \by_j\}}\bigg{|} \by\right]\right]\cr
    &=1-\E_{F_\mcy}\left[\frac{1}{|E|}\sum_{\{i,j\} \in E}\PP(\bx_i\geq \by_i,\bx_j\geq \by_j|\by)\right]\cr
    &=1-\frac{X}{2Y}+\E_{F_\mcy}\left[\frac{X}{4Y^2}\sum_{\{i,j\} \in E}\max\left\{\frac{\by_i}{d_i},\frac{\by_j}{d_j}\right\}\right]\cr
    &\leq1-\frac{X}{2Y}+\E_{F_\mcy}\left[\frac{X}{4Y^2}\sum_{\{i,j\} \in E}\left(\frac{\by_i}{d_i}+\frac{\by_j}{d_j}\right)\right]\cr
    &=1-\frac{X}{2Y}+\E_{F_\mcy}\left[\frac{X}{4Y^2}\sum_{i=1}^n {\by_i}\right] \cr
    &=1-\frac{X}{4Y}.
\end{align*}
\qed

\subsection{Security bounds for general graphs}\label{sec:bounds_proof}

Next, we provide the proof for the lower and upper bounds on the defender's security value $S_\mcx^*$ detailed in Theorem \ref{thm:BoundsGeneral}.

{\it Proof of Theorem \ref{thm:BoundsGeneral}:}
    To establish the lower bound $\gamma(X,Y) \leq S_\mcx^*(X,Y,G)$ \eqref{eq:maxmin_bound}, consider an arbitrary graph $G$ and suppose player $\mcx$ allocates according to the strategy \eqref{eqn:thmFX1} if $X\geq 2Y$, and according to the strategy \eqref{eqn:thmFX2} if $X<2Y$. For $X\geq 2Y$,  we have
    \begin{align*}
        \max_{F_\mcx \in \F(X)}\min _{F_\mcy \in \F(Y)}\pi_\mcx(F_\mcx,F_\mcy)&\geq
        \min _{F_\mcy \in \F(Y)}\pi_\mcx(F^*_\mcx,F_\mcy)\cr
        &=\min_{F_\mcy \in \F(Y)} 1- \pi_\mcy(F^*_\mcx,F_\mcy)\cr
        &\geq 1-\frac{Y}{X}=\gamma(X,Y),
    \end{align*}
    where the last inequality follows the similar argument to the proof of Theorem \ref{thm:NL2}. The proof for the case $X<2Y$ is similar.
    
    To establish the upper bound $S_\mcx^*(X,Y,G) \leq \gamma_n(X,Y)$ \eqref{eq:maxmin_bound}, we consider the following correlated allocation strategy $G_\mcy$ for player $\mcy$. $\mcd$ is the collection of $n$ vertex covers $\{V_k\}_{k\in[n]}$, where we define the $k$th vertex cover as $V_k = V\backslash\{k\} \in \mcd$ for each $k \in V$. In words, the $k$th vertex cover $V_k$ is the set of all nodes except node $k$. We assume each vertex cover is chosen with equal probability $1/n$.
    For simplicity, let $w_{V_k,i}\triangleq w_{k,i}$.
    We set the weights $w_{k,i}$ as follows. For any $k \in V$ and $i \in V\backslash\{k\}$,
    \begin{equation}\label{eq:UB_weights}
        w_{k,i} = 
        \begin{cases}
            \frac{d_i}{2|E|}, &\text{if } k \notin \mcal{N}_i \\
            \frac{d_i+1}{2|E|}, &\text{if } k \in \mcal{N}_i
        \end{cases},
    \end{equation}
    where $ \mcal{N}_i$ is the neighbor set of node $i$ for $i\in V$.
    In words, the weight of node $i$ in vertex cover $k$ is determined by how many nodes it ``covers". For each neighbor already in $V_k$, node $i$ accumulates a weight $\frac{1}{2|E|}$ (they do not need to be covered). If $k$ is a neighbor, then $i$ accumulates a weight of $\frac{1}{|E|}$. It then follows that the sum of weights in any vertex cover is normalized, i.e. $w_{k} \triangleq \sum_{i\neq k} w_{k,i} = 1$. An example illustration of this strategy was provided in Figure \ref{fig:UB_strategy}.
    
    From \eqref{eq:correlated_strat}, the strategy $G_\mcy$ is explicitly written as
    \begin{equation}\label{eq:Y_Kn_stronger}
        G_\mcy(y) = 1-\delta + \frac{\delta^2}{2Yn}\sum_{k=1}^n \min\left\{ \left\{\frac{y_i}{w_{k,i}} \right\}_{i\in V_k}, \frac{2Y}{\delta} \right\},
    \end{equation}
    where $\delta \in [0,1]$ is a tunable parameter. The bi-variate marginal distribution over $i,j \in V$ is then given by
    \begin{equation}
        \begin{aligned}
            G_{\mcy,\{i,j\}}&\left(y_{\{i,j\}}\right) = \\
            &1-\delta + \frac{\delta^2}{2Yn}\Bigg{(}  \min\left\{ \frac{y_i}{w_{j,i}}, \frac{2Y}{\delta} \right\} 
             + \min\left\{ \frac{y_j}{w_{i,j}}, \frac{2Y}{\delta} \right\} \\
             &\hspace{2.5cm}+\sum_{k\neq i,j} \min\left\{ \frac{y_i}{w_{k,i}},\frac{y_j}{w_{k,j}}, \frac{2Y}{\delta} \right\}
             \Bigg{)}.
        \end{aligned}
    \end{equation}
    For any $F_X \in \F(X)$ with the property of Lemma \ref{lem:support_exceed}, i.e. $\mbb{P}_{F_\mcx}(\bx_i > \frac{2Y}{\delta|E|}) = 0$, the expected payoff $\pi_\mcx(F_\mcx,G_\mcy)$ to $\mcx$ is
    \begin{equation}
        \begin{aligned}
            \pi_\mcx(F_\mcx,G_\mcy)&=\frac{1}{|E|}\E_{F_\mcx}\left[ \sum_{\{i,j\} \in E} G_{\mcy,\{i,j\}}(\bx_{\{i,j\}}) \right] \\
            % &= 1-\delta \\
            % &+ \frac{\delta^2}{2Yn} \E_{F_\mcx}\!\!\left[\sum_{\{i,j\}\in E} \!\! \left( \sum_{k \neq i,j} \min\{\frac{x_i}{w_{k,i}},\frac{x_j}{w_{k,j}}\} + \frac{x_i}{w_{k,i}} + \frac{x_j}{w_{k,j}} \right) \right] \\
            &= 1-\delta + \frac{\delta^2}{2Yn|E|}\E_{F_\mcx}\left[ A_1(\bx)+A_2(\bx) \right],
        \end{aligned}
    \end{equation}
    where
    \begin{align}\label{eq:UB_bracket}
        A_1(x)&\triangleq\sum_{\{i,j\}\in E}\sum_{k \neq i,j} \min\left\{\frac{x_i}{w_{k,i}},\frac{x_j}{w_{k,j}} \right\},\cr
        A_2(x)&\triangleq\sum_{\{i,j\}\in E} \left(\frac{x_i}{w_{j,i}}+\frac{x_j}{w_{i,j}}\right).
    \end{align}
    % We can express the term inside the bracket, which we denoted $B(x)$, as
    % \begin{equation}\label{eq:UB_bracket}
    %     \begin{aligned}
    %         &B(x) = \sum_{\{i,j\}\in E}\sum_{k \neq i,j} \min\{\frac{x_i}{w_{k,i}},\frac{x_j}{w_{k,j}} \} + \sum_{i\in V} x_i \left(\sum_{j\in\mcal{N}_i} \frac{1}{w_{j,i}} \right) \\
    %         &\leq \frac{1}{2}\sum_{\{i,j\}\in E}\sum_{k \neq i,j} \left( \frac{x_i}{w_{k,i}} + \frac{x_j}{w_{k,j}} \right) + \sum_{i\in V} x_i \left(\sum_{j\in\mcal{N}_i} \frac{1}{w_{j,i}} \right) \\
    %         &:= A_1(x) + A_2(x)
    %     \end{aligned}
    % \end{equation}
    Due to $\min\{a,b\} \leq \frac{a+b}{2}$, we have
    \begin{equation}
        \begin{aligned}
            A_1(x) &\leq  \frac{1}{2}\sum_{\{i,j\}\in E}\sum_{k \neq i,j} \left( \frac{x_i}{w_{k,i}} + \frac{x_j}{w_{k,j}} \right) \\
            &= \frac{1}{2}\sum_{\{i,j\}\in E}\left( x_i \left( \sum_{k \neq i,j} \frac{1}{w_{k,i}} \right) + x_j \left( \sum_{k \neq i,j} \frac{1}{w_{k,j}} \right) \right), 
        \end{aligned}
    \end{equation}
    For each $\{i,j\} \in E$, we calculate 
    \begin{equation}
        \begin{aligned}
            \sum_{k \neq i,j} \frac{1}{w_{k,i}} &= \sum_{ k \in \mcal{N}_i\backslash \{j\}} \frac{1}{w_{k,i}} + \sum_{k \notin \mcal{N}_i} \frac{1}{w_{k,i}} \\
            &= \frac{2|E|(d_i-1)}{d_i+1} + \frac{2|E|(n-(d_i+1))}{d_i} \\
            &= 2|E|\left(\frac{n-1}{d_i} - \frac{2}{d_i+1}\right)\triangleq \Gamma_i,
        \end{aligned}
    \end{equation}
    where we have used \eqref{eq:UB_weights}. We then obtain
    \begin{equation}
        \begin{aligned}
            A_1(x) &=   \sum_{\{i,j\}\in E}  \left(x_i \Gamma_i + x_j \Gamma_j \right) \\
            &= |E|\sum_{i\in V} x_i d_i\left(\frac{n-1}{d_i} - \frac{2}{d_i+1}\right),
        \end{aligned}
    \end{equation}
    We can also express the second term from \eqref{eq:UB_bracket}, 
    \begin{align*}
        A_2(x)&=|E|\sum_{i\in V} x_i \sum_{j\in\mcal{N}_i} \frac{1}{w_{j,i}} \cr
        &=|E|\sum_{i\in V} x_i \frac{2d_i}{d_i+1},
    \end{align*}
    which is derived using \eqref{eq:UB_weights}. We thus obtain the following upper bound on the payoff for player $\mcx$
    \begin{equation}
        \begin{aligned}
            &\pi_\mcx(F_\mcx,G_\mcy) = 1 - \delta + \frac{\delta^2}{2Yn|E|}\E_{F_\mcx}\left[A_1(\bx) + A_2(\bx) \right] \\
            \ \ &\leq 1 - \delta + \frac{\delta^2}{2Yn}\E_{F_\mcx}\!\!\left[\sum_{i\in V} \bx_i d_i \!\!\left( \frac{n-1}{d_i} - \frac{2}{d_i+1} + \frac{2}{d_i+1}\right) \right] \\
            &= 1 - \delta + \frac{\delta^2(n-1)}{2Yn}\E_{F_\mcx}\left[\sum_{i \in V} \bx_i \right] \\
            &= 1 - \delta + \frac{\delta^2(n-1)}{2n}\frac{X}{Y}.
        \end{aligned}
    \end{equation}
    Recall that the variable $\delta \in [0,1]$ is a parameter that player $\mcy$ can tune. The optimal $\delta^*$ that minimizes the upper bound above is given by
    \begin{equation}
        \delta^* = 
        \begin{cases}
            \frac{nY}{(n-1)X}, &\text{if } X \geq \frac{n}{n-1}Y \\
            1, &\text{if } X < \frac{n}{n-1}Y 
        \end{cases}.
    \end{equation}
    Therefore, player $\mcy$, using $F_\mcy$, can ensure player $\mcx$ obtains a payoff no greater than
    \begin{equation}
        \gamma_n(X,Y,G) =
        \begin{cases}
            1 - \frac{nY}{2(n-1)X}, &\text{if } X \geq \frac{n}{n-1}Y \\
            \frac{n-1}{2n}\frac{X}{Y}, &\text{if } X < \frac{n}{n-1}Y 
        \end{cases}.
    \end{equation}

\qed

\section{Performance of deterministic strategies}\label{sec:sims}

In this section, we highlight the importance of randomized allocation strategies by considering scenarios where the defender ($\mcx$) does not have the ability to randomize its allocation, i.e. it must select a deterministic strategy 
\begin{equation}
    x \in \Delta(X) \triangleq \left\{ x' \in \R_+^n : \sum_{i=1}^n x_i' \leq X \right\}.
\end{equation}
Recall the performance lower bound $\gamma(X,Y)$ from Theorem \ref{thm:BoundsGeneral} is achieved through the randomized strategy \eqref{eqn:thmFX1}. While this lower bound has no explicit dependence on the network structure $G$, we will highlight the dependence of payoff guarantees on network structure when restricted to deterministic strategies through numerical studies.  

We first present some general properties regarding conditions on graph structure and opponent budget $Y$ for which $\mcx$ can guarantee a positive payoff. We then define a heuristic class of deterministic strategies for $\mcx$ and evaluate its performance via numerical simulations on random graphs. These studies not only highlight the the importance of randomized allocations and the impact of different graph structures on performance. 

\subsection{General properties of deterministic strategies}

The following result illustrates some general properties regarding deterministic strategies for $\mcx$ on structured networks.

\begin{proposition}\label{prop:deterministicstrategies}
Let 
\begin{align*}
    S_\mcx^d(X,Y,G)\triangleq\max_{x \in \Delta(X)}\min _{y \in \Delta(Y)}\pi_\mcx(x,y;G)
\end{align*}
be the security value for $\mcx$ under deterministic strategies on any graph $G$. We have
\begin{enumerate}[(a)]
\item  If $G$ is bipartite, $S_\mcx^d(X,Y,G)>0$ iff $Y\leq\frac{X}{2}$.  
    \item If $G$ is a complete graph, $S_\mcx^d(X,Y,G)>0$ iff ${Y\leq\frac{n-1}{n}X}$.
    \item $S_\mcx^d(X,Y,G)=0$ if $Y>\frac{n-1}{n}X$.
    \item $S_\mcx^d(X,Y,G)>0$ if $Y\leq X/2$.
    
\end{enumerate}
\end{proposition}
% Property (a) states that on any graph, $\mcx$ cannot ensure a positive payoff with a deterministic strategy if $Y > \frac{n-1}{n}X$. It can ensure a positive payoff if $Y \leq X/2$. Property (b) states that $Y \leq X/2$ is a necessary and sufficient condition for a positive payoff on bipartite graphs. Property (c) states that $Y \leq \frac{n-1}{n}X$ is a necessary and sufficient condition for a positive payoff on complete graphs.
Proposition \ref{prop:deterministicstrategies}  identifies bipartite and complete graphs as the two extreme network structures determining a defender's effectiveness. Indeed, bipartite networks provide positive payoff guarantees on the smallest range of parameters, i.e. $Y \in [0,X/2]$ (property (a)). Complete networks provide positive payoff guarantees on the widest range of parameters, i.e. for any $Y \in [0,\frac{n-1}{n}X]$ (property (b)).  The range of parameters where this guarantee holds for an arbitrary graph $G$ must lie somewhere in between, i.e. $Y \in [0,fX]$ for some $\frac{1}{2} \leq f \leq \frac{n-1}{n}$ (property (c)-(d)). The proof of the Proposition is provided in  Section \ref{sec:proof}.

\begin{proof}
    First, we prove parts (c) and (d).
    
     To prove part (c), let set $A_i=[n]\setminus\{i\}$ for $i\in[n]$. First, note that $A_i$s are vertex cover. Therefore, if for some $i\in[n]$,
    % \begin{align*}
        $Y>X_i\triangleq \sum_{j\in A_i}x_j,$
    % \end{align*}
     assigning ${y_j=x_j+\frac{Y-X_i}{n-1}}$ for all $j\in A_i$, player $\mcy$ wins all vertices in $A_i$, and hence, wins all the edges. On the other hand, We have
    \begin{align*}
        \sum_{i=1}^n\sum_{j\in A_i}x_j=\sum_{j=1}^n(n-1)x_j=(n-1)X.
    \end{align*}
    This implies
    \begin{align*}
        \min_{j\in[n]}\left\{\sum_{j\in A_j}x_j\right\}\leq \frac{n-1}{n}X,
    \end{align*}
    which completes the proof of $S_\mcx(X,Y,G)=0$ if $Y>\frac{n-1}{n}X$.
    
    To prove part (d), suppose that there is an edge between nodes $i$ and $j$, and $x_i=x_j=\frac{X}{2}$. Therefore, in order for player $\mcy$ to win this edge, $Y$ should be strictly greater $\frac{X}{2}$. 
    
    For part (a), note that ``if" part is implied by part (d). To prove the the other direction, let $B_1$ and $B_2$ be the vertex set of two part of the a bipartite graph $G$, i.e., $B_1\cup B_2=V$, and there is no edge between any two vertices in $B_1$ or $B_2$. Since
    % \begin{align*}
        $\sum_{j\in B_1}x_j+\sum_{j\in B_2}x_j=X$,
    % \end{align*}
    we have 
    \begin{align*}
        \min\left\{\sum_{j\in B_1}x_j,\sum_{j\in B_2}x_j\right\}\leq \frac{X}{2}.
    \end{align*}
    Therefore, if $Y>\frac{X}{2}$, player $\mcy$ can win all vertices in $B_1$ or $B_2$, and hence, win all the edges.
    
    For part (b), note that ``only if" part is implied by part (c). To prove the the other direction, we note that every vertex cover in a complete graph has $n-1$ vertices. If player $\mcx$ assign his budget uniformly, i.e., $x_i=\frac{X}{n}$ for $i\in[n]$, $Y$ has to be strictly greater than $\frac{n-1}{n}X$ in order for player $\mcy$ to win all edges. 

\end{proof}

In the following, we will quantify and compare the performance of a certain class of deterministic strategies on various networks through simulations. These experiments corroborate the findings of Proposition \ref{prop:deterministicstrategies}.

\subsection{Numerical simulations}

On arbitrary networks, we evaluate the performance of the following deterministic strategy for $\mcx$:
\begin{equation}
    x_i^{(d)} \triangleq \frac{d_i}{2|E|} X, \quad \forall i \in V,
\end{equation}
where $d_i$ is the degree of node $i$.
In words, the amount of resources $\mcx$ allocates to any node is proportional to its degree centrality\footnote{This deterministic strategy is inspired by its randomized counterpart \eqref{eqn:thmFX1}, which is optimal  for bipartite graphs.}.

% $x^{(r)}$ parameterized by a number $r \in (0,1]$. Given a graph $G$, denote $V_r$ as the set of $\lfloor rn \rfloor$ nodes with the highest degree centralities. Then, we define the strategy $x^{(r)} \in \Delta(X)$ as the allocation to nodes in $V_r$ proportional to their degree:
% \begin{equation}\label{eq:xr}
%     x_i^{(r)} \triangleq 
%     \begin{cases}
%         \frac{d_i}{\sum_{j \in V_r} d_j} X, &\text{if } i \in V_r \\
%         0, &\text{if } i \notin V_r
%     \end{cases}
% \end{equation}

In response, player $\mcy$ then selects a \emph{deterministic} strategy $y \in \Delta(Y)$. The best-response problem for $\mcy$, i.e $\max_{y\in\Delta(Y)} \pi_\mcy(x,y)$ for any $x \in \Delta(X)$ is known as a \emph{budgeted maximum coverage} problem, which is NP-hard. To evaluate the performance of a defender that implements $x^{(d)}$, we adapt a greedy algorithm from \cite{Khuller_1999} to approximate $\mcy$'s best-response, which we detail in Algorithm 1. This gives a performance factor\footnote{Other methods can improve the guarantee to $1 - 1/e$ \cite{Khuller_1999}. We do not utilize these, however, because they come with a much higher computational cost than Algorithm 1, especially on large scale networks.} of $1 - 1/\sqrt{e}$. 

% Briefly, the greedy response sequentially secures nodes with the highest degree-to-cost ratio $(\EE(N \cup i) - \EE(N))/x_i$, where $N \subset V$ is an accumulated subset of nodes chosen from previous iterations, and $\EE(N) \triangleq |\{(i,j) \in E : i \text{ or } j \in N\}|$ is the number of edges connected to nodes in $N$. A final comparison is made between the edges accumulated through this procedure and the edges taken from any one single node (line 12).

% Heuristic procedures can approximate the optimal response within a certain guaranteed factor from the optimal.

% In summary, the greedy response $y(x)$ generated by Algorithm 1 to any $x \in \Delta(X)$ is determined such that $\mcy$ sequentially secures nodes with the highest degree-to-cost ratio $(\EE(N \cup i) - \EE(N))/x_i$, where $N \subset V$ is an accumulated subset of nodes chosen from previous iterations, and $\EE(N) \triangleq |\{(i,j) \in E : i \text{ or } j \in N\}|$ is the number of edges connected to nodes in $N$. In the case that there are ties, a random draw from the set of cheapest nodes is chosen. This naive procedure alone has arbitrarily bad performance guarantees \cite{Khuller_1999,Nguyen_2013}. A final comparison is thus made between the edges accumulated through this naive procedure and the edges taken from any one single node (line 12). This slight modification guarantees performance within a $1 - 1/\sqrt{e}$ factor of the optimal \cite{Khuller_1999}.

% This univariate distribution is a best-response to the pure strategy $x_i$ in a standard Lotto game with a single battlefield.

\begin{figure}
    \centering
    \includegraphics[scale=0.35]{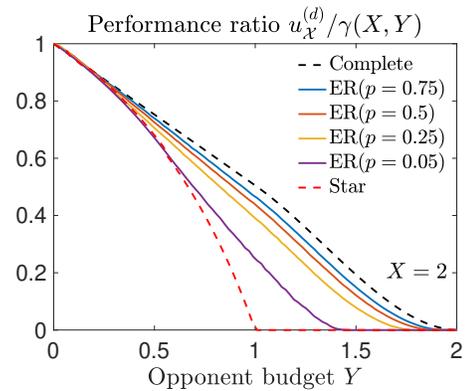}
    \caption{Simulations showing the  deterministic-to-randomized performance ratio $u_\mcx^{(d)}/\gamma(X,Y)$ on a variety of networks. We fix $n=100$ for all graphs considered. For ER random graphs, we average the performance over 100 independent samples at each value of $Y$. Here, $p$ is the independent link formation probability. Deterministic strategies perform better on more densely connected networks. In all simulations, we fix $X=2$.}
    
    % Here, $\gamma$ \eqref{eq:piXstar} is a benchmark lower bound performance for $\mcx$ using randomized strategies over arbitrary networks (Theorem \ref{thm:BoundsGeneral}).
    
    % Against $x^{(r)}$, player $\mcy$ generates a greedy response according to Algorithm 1. These plots thus serve as upper bounds on $\mcx$'s performance in relation to actual best-responses from $\mcy$ \eqref{eq:numeric_perf_bounds}. 
    
    % (Left) The ratio $u_\mcx^{(r)}/\pi_X^*$ for varying values of $r \in (0,1]$, averaged over 100 independent samples of ER graphs with $n=100$ and $p = 0.25$ for each value of $Y$. The ratio is monotonically increasing in $r$.

    \label{fig:det_sims}
\end{figure}

% Prove there are no pure equilibria in Lotto network game
% Network dependence
% Y best-response to x is NP hard. 

We denote the resulting payoff to $\mcx$ as $u_\mcx^{(d)} \triangleq \pi_\mcx(x^{(d)},y(x^{(d)}))$, where $y(x^{(d)}) \in \Delta(Y)$ is the response generated from Algorithm 1.  Figure \ref{fig:det_sims} shows numerical simulations characterizing the performance ratio $u_\mcx^{(d)}/\gamma(X,Y)$ on Erd{\"o}s-R{\'e}nyi (ER) random networks, where $\gamma(X,Y)$ is the lower bound \eqref{eq:piXstar} on the payoff that $\mcx$'s can ensure on any graph using randomized strategies (Theorem \ref{thm:BoundsGeneral}). Hence, $\gamma(X,Y)$ serves as a benchmark comparison. 

There are two main conclusions from these experiments. First, $x^{(d)}$ performs better on graphs with higher edge density. The performance is worst on star, ring, and line networks, improves on ER random networks for higher link parameter $p \in [0,1]$, and performs best on the complete network (Figure \ref{fig:det_sims}, right). These numerical findings are consistent with the results established in Proposition \ref{prop:deterministicstrategies}. Second, deterministic strategies perform significantly worse relative to randomized strategies. Considering that we use the lower bound $\gamma(X,Y)$ as the benchmark comparison, and the greedy response is sub-optimal for $\mcy$, the performance ratio $u_\mcx^{(d)}/\gamma(X,Y)$ serves as an upper bound on the performance of $x^{(d)}$. Specifically, it is an upper bound on the ratio wherein $\mcy$ implements a best response and the denominator is taken as the actual max-min value for $\mcx$ (LHS of \eqref{eq:maxmin_bound}). 

% First, the best performing deterministic strategy is $x^{(1)}$, i.e. allocating proportionally to \emph{every} node in the network based on its degree (Figure \ref{fig:det_sims}, left).

\begin{remark}
    The strategy $x^{(d)}$ is a heuristic strategy for $\mcx$ used for the purpose of making performance comparisons on arbitrary networks via numerical simulations. It is not proven here that $x^{(d)}$ is optimal among deterministic strategies, i.e. one that solves $\max_{x\in\Delta(X)} \min_{y \in \Delta(Y)} \pi_\mcx(x,y)$. Such a problem is difficult to solve, since the inner minimization alone is NP-hard. 
    
    % Nevertheless, our numerical simulations provide an upper bound on the performance of the strategy $x^{(r)}$:
    % \begin{equation}\label{eq:numeric_perf_bounds}
    %     \begin{aligned}
    %         u_\mcx^{(r)} &\geq \min_{y \in \Delta(Y)}  \pi_\mcx(x^{(r)},y) \\ 
    %         &\geq \min_{F_\mcy \in \F(Y)} \pi_\mcx(x^{(r)},F_\mcy)
    %     \end{aligned}
    % \end{equation}
    
    % Moreover, the greedy algorithm from \cite{Khuller_1999} gives the performance factor $1 - 1/\sqrt{e}$ in \emph{general} budgeted maximum coverage problems, i.e. ones with arbitrary submodular evaluation functions, set elements, and costs. Our setup is a special case, where the costs are given precisely by \eqref{eq:xr}, and the set elements are edges of a network. It would be of interest to show whether there exist efficient algorithms that provide approximation factors better than $1 - 1/e$ in these specific cases.
\end{remark}

% \begin{algorithm}
% \caption{Greedy response $F_\mcy^x$ to $x$ on a network $G$}\label{alg:algorithm}
% \begin{algorithmic}[1]
% \Procedure{greedy}{$x,Y,G$} %\Comment{The g.c.d. of a and b}
% \State $F_{\mcy,i}^x \triangleq \delta_{0}$ \ $\forall i \in V$ \Comment{initialize no allocations}
% \State $V_0 \triangleq \{i \in V : x_i = 0\}$ \Comment{nodes w/ no $x$ resources}
% \State $E_\mcy \triangleq \{(i,j) \in E : i \text{ or } j\in V_0\}$ \Comment{$\mcy$-secured edges}
% % \State $\NL_i(E_\mcy) \triangleq \{ (i,j)\in E\setminus E_\mcy \}$ \ $\forall i \in V$

% \While{$Y \geq \max_{i \in V} |\EE_i(E_\mcy)|$}
%     \State $i^* \gets \arg\max_{i\in V} |\EE_i(E_\mcy)|$
%     \State $F_{\mcy,i^*}^x \gets \delta_{x_{i^*}}$ \Comment{Allocation secures $i^*$}
%     \State $Y \gets Y - x_{i^*}$ \Comment{Update remaining budget}
%     \State $E_\mcy \gets E_\mcy \cup \EE_{i^*}(E_\mcy)$ \Comment{$\mcy$-secured edges}
% \EndWhile
% \State $i^* \gets \arg\max_{i\in V} |\EE_i(E_\mcy)|$
% \State $F_{\mcy,i^*}^x \gets (1-\frac{Y}{x_{i^*}})\delta_0 + \frac{Y}{x_{i^*}}\delta_{x_{i^*}}$ \Comment{Randomize on last node}
% \State $U_Y \triangleq |E_\mcy| + \frac{Y}{x_{i^*}}|\EE_{i^*}(E_\mcy)|$  \Comment{Resulting payoff}
% \State \textbf{return} $F_\mcy^x, U_Y$ \Comment{Greedy response and its payoff}
% \EndProcedure
% \end{algorithmic}
% \end{algorithm}

\begin{algorithm}
\caption{Greedy response of $\mcy$ to $x \in \Delta(X)$ on $G = (V,E)$. This algorithm is adapted from \cite{Khuller_1999}. We define $\EE(N) \triangleq |\{(i,j) \in E : i \text{ or } j \in N\}|$ as the number of edges connected to the subset of nodes $N \subset V$ (line 7). }\label{alg:algorithm}
\begin{algorithmic}[1]
\Procedure{greedy}{$x,Y,G$} %\Comment{The g.c.d. of a and b}
\State $y_i \gets 0$, $\forall i \in V$ \Comment{initialize response}
\State $i^* \gets \max_{i\in V} \EE(i)$ \Comment{highest degree node}
\State $N \gets \varnothing$ \Comment{nodes to secure}
% \State $N \gets \{i \in V : x_i = 0\}$ \Comment{nodes w/ zero cost}
% \State $V \gets V \setminus N$
\While{$V \neq \varnothing$}
    \State $D = \arg\max_{i \in V} \{ \frac{\EE(N\cup i) - \EE(N)}{x_i} \}$
    \State $u \gets \arg\min_{v \in D} x_v$ \Comment{cheapest node in $D$}
    \If{$\sum_{j \in N\cup u} x_j \leq Y$} \Comment{budget-feasible}
        \State $N \gets N \cup u$ \Comment{update secured nodes}
    \EndIf
    \State $V \gets V \setminus u$
\EndWhile
\State $\EE^* \gets \max\left\{1(x_{i^*} \leq Y)\cdot\EE(i^*), \EE(N) \right\}$
\If{$\EE^* = \EE(N)$}
    \State $y_i \gets x_i$, $\forall i \in N$
\Else
    \State $y_{i^*} \gets x_{i^*}$
\EndIf 
\State \textbf{return} $y \in \Delta(Y)$ \Comment{greedy response}
\EndProcedure
\end{algorithmic}
\end{algorithm}

\section{Conclusion and Future Research}\label{Conclusion}

In the context of General Lotto games, we studied a competitive resource allocation scenario between an attacker and defender of a network. The objective for each player is to secure as many edges of the network as possible, where the defender is required to win both endpoint nodes in order to secure an edge. We completely characterized equilibrium payoffs and strategies for both players when the network is bipartite. On arbitrary networks, we provided lower and upper bounds on the defender's max-min performance. 
% Numerical methods are able to calculate tighter upper bounds on arbitrary networks. 
To further demonstrate the impact of network topology, we then considered a defender restricted to deterministic strategies. We identified bipartite and complete graphs as the two extreme structures that determine the defender's effectiveness. These findings were corroborated through numerical simulations. 

% Understanding which network topologies are easier to defend and harder to attack will provide sharper tools to inform more robust system designs. Characterizations of the optimal security resource allocations  over networks provides principled guidelines for defending against strategic adversaries.

It is of interest to extend our equilibrium results to larger classes of networks. In Section \ref{sec:sims}, we saw that the performance of deterministic strategies is highly dependent on the network structure. We would like to identify the salient network characteristics, e.g. edge density, cluster coefficients, etc, that contribute to the performance. 

% and the attacker's best-response problem is NP-hard. It would be of interest to analyze or devise algorithms that provide sharper performance guarantees

\bibliographystyle{abbrv}
\bibliography{source}

\begin{IEEEbiography}[{\includegraphics[width=1in,height=1.25in,clip,keepaspectratio]{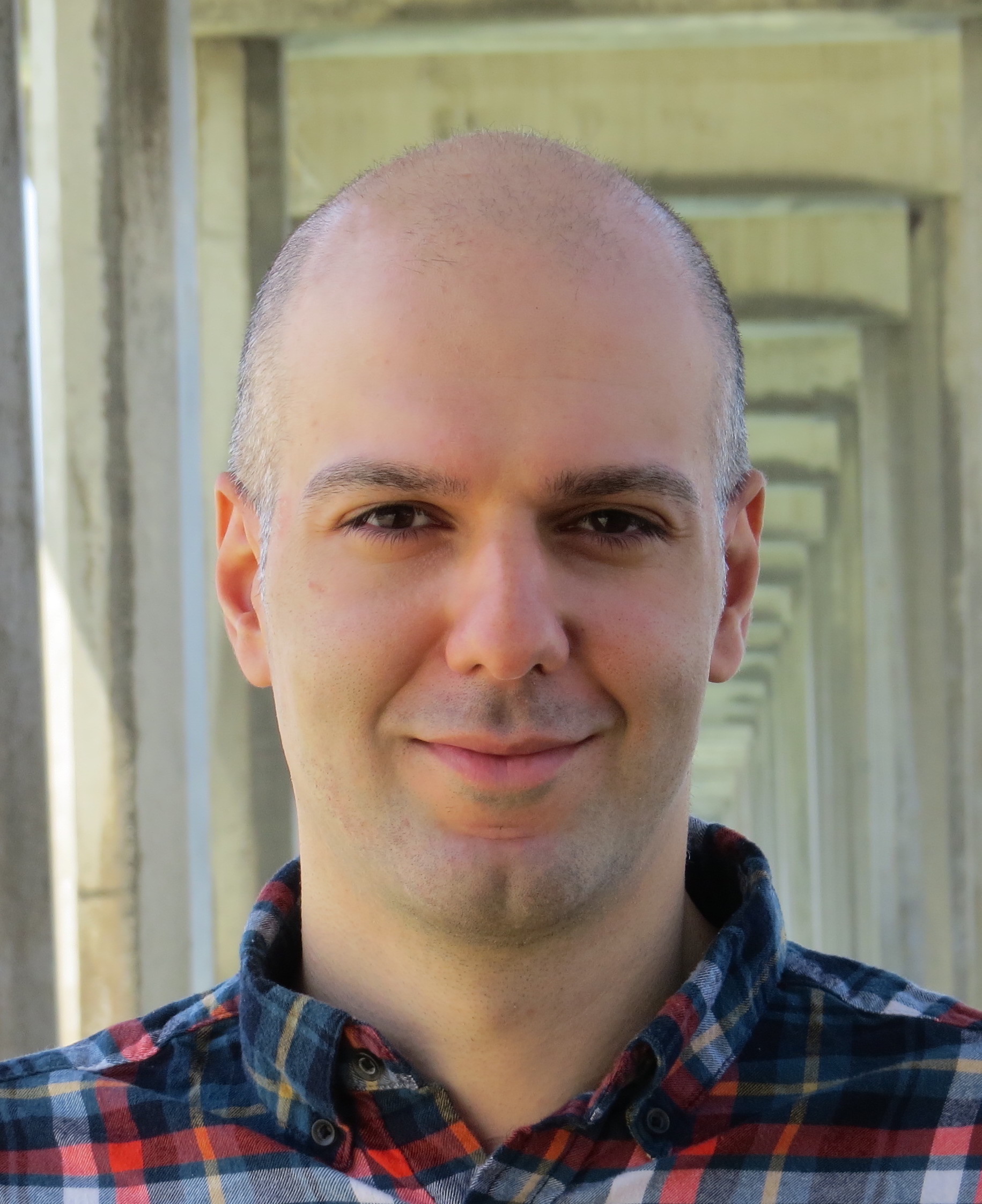}}] {Adel Aghajan}  received a B.Sc., and M.Sc. degrees in Electrical Engineering
 from Isfahan University of Technology, Isfahan, Iran, in 2009, and 2012,
respectively, and  a Ph.D.\ degree in Electrical Engineering at University of California, San Diego, in 2021. He is currently a Postdoctoral Scholar with the
Department of Electrical and Computer Engineering, the University of California, Santa Barbara. His research interests include game theory, distributed computation
and optimization
, random dynamics, and information theory. 
    
\end{IEEEbiography}

\begin{IEEEbiography}[{\includegraphics[width=1in,height=1.25in,clip,keepaspectratio]{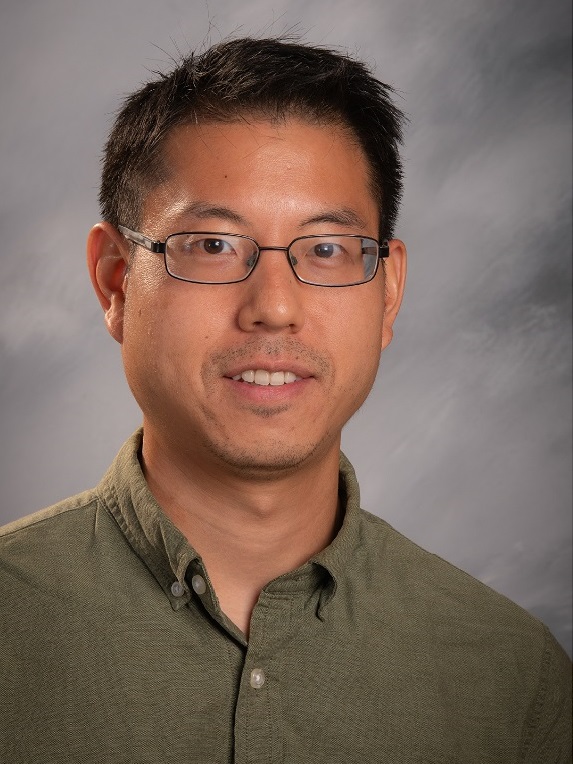}}] {Keith Paarporn}
  is an Assistant Professor in the Department of Computer Science at the University of Colorado, Colorado Springs. He received a B.S. in Electrical Engineering from the University of Maryland, College Park in 2013, an M.S. in Electrical and Computer Engineering from the Georgia Institute of Technology in 2016, and a Ph.D. in Electrical and Computer Engineering from the Georgia Institute of Technology in 2018. From 2018 to 2022, he was a postdoctoral scholar in the Electrical and Computer Engineering Department at the University of California, Santa Barbara. His research interests include game theory, control theory, and their applications to multi-agent systems and security. 
\end{IEEEbiography}

\begin{IEEEbiography}[{\includegraphics[width=1in,height=1.25in,clip,keepaspectratio]{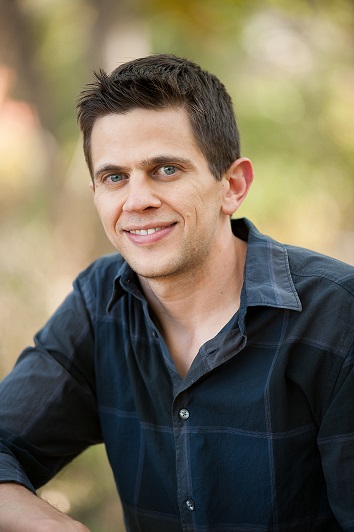}}] {Jason Marden}
    is a Professor in the Department of Electrical and Computer Engineering at
    the University of California, Santa Barbara. Jason received a BS in Mechanical Engineering in 2001 from
    UCLA, and a PhD in Mechanical Engineering in
    2007, also from UCLA, under the supervision of Jeff
    S. Shamma, where he was awarded the Outstanding
    Graduating PhD Student in Mechanical Engineering.
    After graduating from UCLA, he served as a junior
    fellow in the Social and Information Sciences Laboratory at the California Institute of Technology until
    2010 when he joined the University of Colorado. Jason is a recipient of the NSF Career Award (2014), the ONR Young Investigator Award (2015), the AFOSR Young Investigator Award (2012), the American Automatic Control Council Donald P. Eckman Award (2012), and the SIAG/CST Best SICON Paper Prize (2015). Jason’s research interests focus on game theoretic methods for the control of distributed multiagent systems.
\end{IEEEbiography}

\end{document}